\documentclass[journal,12pt,onecolumn]{IEEEtran}
\usepackage{fixltx2e}[2006/09/13]

\usepackage[round,sort]{natbib}
\usepackage{multirow}

\ifCLASSINFOpdf
   \usepackage[pdftex]{graphicx}
   \graphicspath{{images/}}
   \DeclareGraphicsExtensions{.pdf,.jpeg,.png}
\else
\fi

\usepackage[cmex10]{amsmath}
\usepackage{amsfonts}
\usepackage{amsthm}
\usepackage[caption=false]{subfig}
\usepackage{epstopdf}
\usepackage{booktabs}
\usepackage{algorithm}
\usepackage{algpseudocode}
\usepackage[nolist]{acronym}
\usepackage{tikz}

\newtheorem{proposition}{Proposition}
\newtheorem{example}{Example}

\begin{document}
%
\title{The Membership Degree Min-Max\\ Localisation Algorithm}

\author{
\IEEEauthorblockN{Thomas Hillebrandt, Heiko Will, Marcel Kyas}
\IEEEauthorblockA{
  \\Freie Universit\"at Berlin\\
  AG Computer Systems \& Telematics\\
  Berlin, Germany\\
  Email: \{thomas.hillebrandt,heiko.will,marcel.kyas\}@fu-berlin.de}\\
  Telephone: +49 30 838 75116 \\
  Fax: +49 30 838 75194 \\ 
}


\maketitle
\tikz[remember picture,overlay] {
    \node[shift={(10.7cm,1.0cm)}] at (current page.south west) {\footnotesize \hfil\parbox[t][\height][t]{\textwidth}{\centering \copyright~[2012] IEEE. Portions reprinted, with permission, from Will H., Hillebrandt T., Yang Y., Zhao Y., Kyas M.,\\ The Membership Degree Min-Max localization algorithm, IEEE Proceedings of UPINLBS 2012, October/2012.}\hfil\hbox{}
};
}

\begin{abstract}

We introduce the \ac{MD-Min-Max} localisation algorithm as a precise and simple lateration algorithm for indoor localisation. \ac{MD-Min-Max} is based on the well known Min-Max algorithm that computes a bounding box to estimate the position. \ac{MD-Min-Max} uses a \ac{MF} based on an estimated error distribution of the distance measurements to improve the precision of Min-Max. The algorithm has similar complexity to Min-Max and can be used for indoor localisation even on small devices, e.g. in \acp{WSN}. To evaluate the performance of the algorithm we compare it with other improvements of the Min-Max algorithm and maximum likelihood estimators, both in simulations and in a large real world deployment of a \ac{WSN}. Results show that \ac{MD-Min-Max} achieves the best performance in terms of average positioning accuracy while keeping computational cost low compared to the other algorithms.
\end{abstract}


\IEEEpeerreviewmaketitle

\newpage
\section{Introduction}

Localising an item by using distance measurements, so called lateration, is not only of interest to computer scientists. Localisation algorithms are used in many other domains. For example, psychologists want to detect the precise spatial source of a electric impulse in the human brain, biologists want to track the position of birds equipped with small sensor nodes and geologists have to detect the source of an earthquake using seismic waves. Most of those applications apply the same principle: based on the measurement of a physical value, the distance between the target and some fixed points (anchors) is estimated and then the position of the target is calculated with a localisation algorithm. Since the distance estimations are erroneous, classic algorithms like trilateration provide unsatisfactory results. Thus, many algorithms have been proposed, that turn out to be robust in different geometrical constellation of target and anchors.

In this paper we present \acl{MD-Min-Max} (\acs{MD-Min-Max}), a localisation algorithm which is based on the well known Min-Max algorithm~\citep{whitehouse2006robustness,lazos2005serloc}. \ac{MD-Min-Max} can easily be adapted to the distance error distribution of a fixed anchor deployment to minimize the position error. To make use of this algorithm the error distribution must be known which should be easy to achieve for static deployments. Especially for indoor localisation of nodes our algorithm shows a big improvement on the position error because it weakens the effect of multi-path propagation and signal reflection. Even for unknown error distributions in dynamic environments our algorithm performs quite well with a general distribution function. We show in several simulations and a real world deployment that the position error is minimal compared to other Min-Max solutions and maximum likelihood estimators while keeping the computation and memory complexity low.

The main difference to other methods is that we do not simply take the centroid of the bounding box, but weight the vertexes of the bounding box with a membership function based on the distance error distribution of the deployment. With this weighting we improve precision even in the areas where other Min-Max algorithms performance gets worse, which is especially outside the convex hull of the anchors.

One property of localisation algorithms is the \emph{spatial distribution of the positioning error}, or shorter, the \emph{spatial error distribution}. This measures the expected positioning error at each location in a space. It characterises the performance of the algorithm for a specific anchor set-up and shows where an algorithm is expected to perform well and where it does not perform. For example, the Min-Max algorithm will only compute positions inside the convex hull that contains all anchor positions. Outside of this hull, the positioning error increases until the algorithm becomes unusable. We analyse the spatial error distribution of our algorithm and compare the results to other algorithms.

The remainder of the paper is structured as follows. In Section~\ref{relatedwork} we present related work and the original Min-Max algorithm. Furthermore, we introduce the \ac{E-Min-Max} \citep{extended2012} algorithm that also uses a weighting function to improve the precision of Min-Max. We also describe \ac{ML} estimation \citep{al2002ml,patwari2003relative} which uses an error model to account for the distance measurement characteristics. As with \ac{MD-Min-Max}, prior knowledge of the probability density is needed and modelling the indoor sensor error is a difficult task. We fit the error distribution obtained by experiments in a real environment using normal and gamma distribution. In Section~\ref{section:md-min-max} we describe our algorithm in detail and discuss how to calculate the membership function. In Section~\ref{section_spatial_evaluation} we present an evaluation of the spatial error distribution of the selected algorithms. After that we discuss the results of our real world deployment in Section~\ref{deployment}. Finally our conclusion and future work is presented in Section~\ref{conclusion}.


\newpage
\section{related work}
\label{relatedwork}

Several measurement techniques are used to track the positions for indoor systems~\citep{patwari2005locating,bahl2000radar,bahl2000enhancements,ranta2011accu}. Range based methods which measure the distance or range value between the target and anchor sensors are common and efficient tools, for instance, received signal strength (RSS) in RADAR system~\citep{bahl2000radar,bahl2000enhancements}, time-of-arrival (TOA) and its improved metrics: time-difference-of-arrival (TDOA)~\citep{patwari2005locating} and \ac{TOF}~\citep{bulusu2000gps}. \ac{TOF} measures the round-trip time of packet and averages the result together to reduce the impact of time-varying errors. It is a promising solution for its low cost and feasible for the capacity of real-time application.

Range based location algorithms are designed to reduce range errors such as the complicated indoor multi-path propagations, low signal-to-noise ratio (SNR), severe multi-path effects, reflection and link failures and improve the estimation accuracy~\citep{whitehouse2005effects,whitehouse2007practical,guvenc2007analysis,venkatraman2004novel,Venkatesh2006}. These algorithms include iterative methods, which use gradient descent or Newton method to calculate an estimated position. Grid-scan methods~\citep{lazos2005serloc,srinivasan2007survey} divide the target field into several cells and are using voting based methods to select a cell as an estimated position. Refined geometry relationship~\citep{venkatraman2004novel,moore2004robust} obtains the target relative position rather than actual position, and the method is still based on the range based measurements, in which the measurement noise still causes estimation errors. Least squares (LS) method~\citep{guvenc2007analysis,Venkatesh2006} can be classified into linear least squares (LLS) algorithm and nonlinear least squares (NLLS) algorithm. LS is a common and accurate way for localisation, however, the achieved solution is suboptimal in case the estimated distances contain outlier errors~\citep{beck2008exact}. Optimal range selection~\citep{kaplan2006global,li2010low} directly reduces the range error by adapting the range measurement and choosing effective anchors.

Most of the common algorithms do not perform very well in indoor scenarios. Indoor scenarios are commonly classified by a large number of anchors with a short inter-anchor distance. The error on the distance measurements is often biased for a subset of the overall anchor configuration due to multi-path effects and reflections of the received signals. The Min-Max~\citep{simic01:_distr} algorithm is an effective and simple method for localisation. Experiments show, that the Min-Max method performs very well in short-range scenarios~\citep{zanca08:_exper_rssi}.

\subsection{Min-Max}

\begin{figure}[htb]
  \centering
  \includegraphics[width=3.5in]{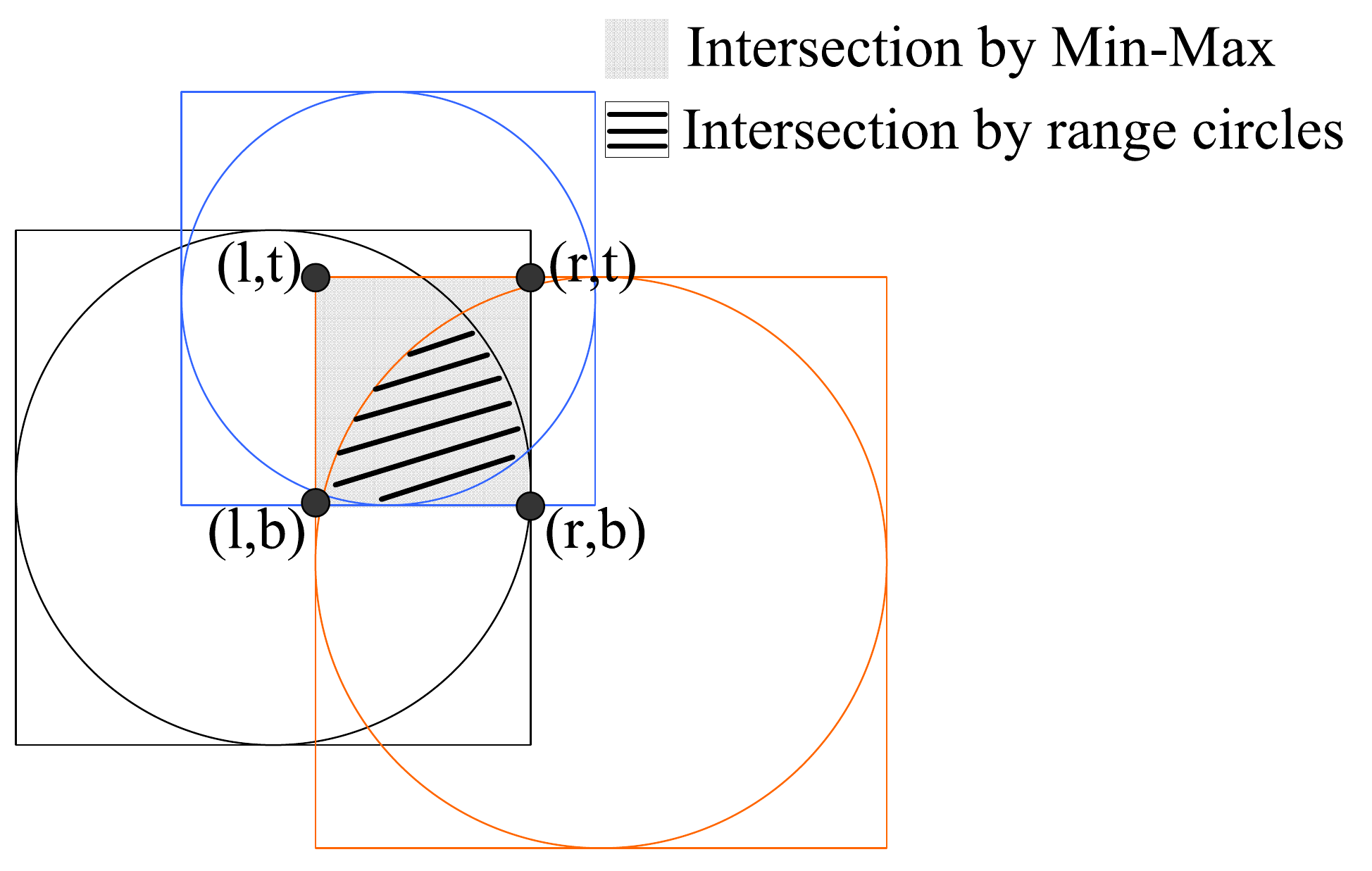}
  \vspace{-0.1cm}
  \caption[The geometrical representation of the intersection by range circles or boxes.]{}
  \label{fig:box}
  \vspace{-0.1cm}
\end{figure}

The Min-Max algorithm, also known as Bounding Box algorithm, is a simple and straightforward method. It contains only very few arithmetic operations, the run-time complexity is in $\Theta(N_{\mathit{anc}})$. Min-Max builds a square (bounding box) given by $[a_{xi}-r_i,a_{yi}-r_i] \times [a_{xi}+r_i,a_{yi}+r_i]$ around each anchor node~$i$ using its location $a_i=(a_{xi},a_{yi})$ and distance estimate $r_i$, instead of using circles with radius $r_i$. The position of target satisfies every box, thus the position is in the \ac{IR} with vertices $\mathbf{V}=\{(l,b),(r,b),(l,t),(r,t)\}$, as Eq.~\eqref{eq:box} and Figure~\ref{fig:box}. Then, estimation of position $(\hat{x},\hat{y})=(\frac{l+r}{2},\frac{t+b}{2})$ is the centre of \ac{IR}.

\begin{equation}
  \label{eq:box}
 \mathit{IR}= \bigcap\limits_{i=1}^{N_{anc}} {\{a_{xi}-r_i , a_{xi}+r_i, a_{yi}-r_i , a_{yi}+r_i\} }~,
\vspace{12pt}
\end{equation}
where $N_{anc}$ denotes the number of anchors and with

\begin{equation}
\label{eq:vertex}
\renewcommand{\arraystretch}{1.3}
\left\{ \begin{array}{l}
\vspace{ 0.17cm}
 l = \max_{i = 1}^{N_{anc}} \{a_{xi}-r_i \} \\
\vspace{ 0.17cm}
 r =\min_{i = 1}^{N_{anc}} \{a_{xi}+r_i \} \\
\vspace{ 0.17cm}
 t =\min_{i = 1}^{N_{anc}} \{a_{yi}+r_i \} \\
\vspace{ 0.0cm}
 b =\max_{i = 1}^{N_{anc}} \{a_{yi}-r_i \}.\\
\end{array} \right.
\vspace{12pt}
\end{equation}

However, Min-Max can produce high position error even when having small distance measurement error, particularly when the target is located outside the perimeter of the anchor nodes. Due to the multi-path effect, most of the measured distances are larger than the actual distance, which is especially common in indoor scenarios. Furthermore, the box has larger area than the corresponding range circle. Even though the range is imprecise, the target is more likely to resist in \ac{IR} or be close to \ac{IR}. Therefore, to find a more reasonable estimation in \ac{IR} can be a potential method to increase the location accuracy.

\subsection{Extended Min-Max}

\ac{E-Min-Max} determines the \ac{IR} the same way Min-Max does but the position of the unlocalised node can be located at any point inside the \ac{IR} and not only at the centre of it. Therefore, \ac{E-Min-Max} assigns a weight \begin{math}W_{a}\end{math} to each vertex of the \ac{IR}. In the original paper \ac{E-Min-Max} is evaluated with four different weights (\begin{math}W_1, W_2, W_3, W_4\end{math})~\citep{extended2012}. We limit our evaluation to the two weights which showed the best performance, \begin{math}W_2\end{math} and \begin{math}W_4\end{math}:
\vspace{12pt}
\begin{equation}
W_{2}(j)= \frac{1}{\sum_{i=1}^n(D_{i,j} - r_i)^2}
\label{e_min_max_w2}
\end{equation}

\begin{equation}
W_{4}(j)= \frac{1}{\sum_{i=1}^n |D_{i,j}^2 - r_i^2|}
\label{e_min_max_w4}
\vspace{12pt}
\end{equation}
where \begin{math}D_{i,j}\end{math} is the Euclidean distance between anchor \begin{math}i\end{math} and vertex \begin{math}j\end{math} of the \ac{IR}. In general, \begin{math}W_4\end{math} gives better results inside the perimeter of the anchors and \begin{math}W_2\end{math} shows the best overall performance, even outside the perimeter of the anchors. The final position is estimated by calculating the weighted centroid with the weights and the coordinates of the vertices as in Eq.~\eqref{e_min_max_pos}.
\vspace{12pt}
\begin{equation}\label{e_min_max_pos}
  (\hat{x},\hat{y})= \left(\frac{\sum_{j=1}^4 W_a(j) \cdot x_j}{\sum_{j=1}^4 W_a(j)}, \frac{\sum_{j=1}^4 W_a(j) \cdot y_j}{\sum_{j=1}^4 W_a(j)}\right)
\end{equation}

Compared to the original Min-Max, \ac{E-Min-Max} requires extra operations to estimate the weights for the vertices. Especially, \ac{E-Min-Max}~(W2) includes square roots which is more expensive in terms of computation but the run-time complexity of \ac{E-Min-Max} is also in $\Theta(N_{\mathit{anc}})$.

Weighting with the absolute residues is based on the assumption that $|D_{i,j} - r_i|$ can approximate $|D_{i,j} -\bar{r}_i|$, where $\bar{r}_i$ is the $i$th actual distance. However, some distance estimation errors are extremely large due to \ac{NLOS} propagation, which results in large residues even if close to the actual target position. Thus, \ac{E-Min-Max} cannot improve the accuracy in some cases and still the error distribution for real environment is not considered.

\subsection{Maximum Likelihood Estimation}

\ac{ML} estimation can be employed if the \ac{pdf} of the distance error contained in the noisy distance measurements is known. Let $p(r\mid u)$ denote the \ac{pdf} that specifies the probability of observing the distance measurements $r$ at the position $u$ of the unlocalised node. \ac{ML} estimation calculates the source location as the value $u$ that maximizes the likelihood function, i.e.,
\vspace{12pt}
\begin{equation}
\hat{u}_{\textit{ML}} = \underset{u}{\operatorname{argmax}} \ p(r\mid u)
\label{likelihood_function_estimate}
\end{equation}

\begin{figure}[htb]
  \centering
  \includegraphics[width=.70\linewidth]{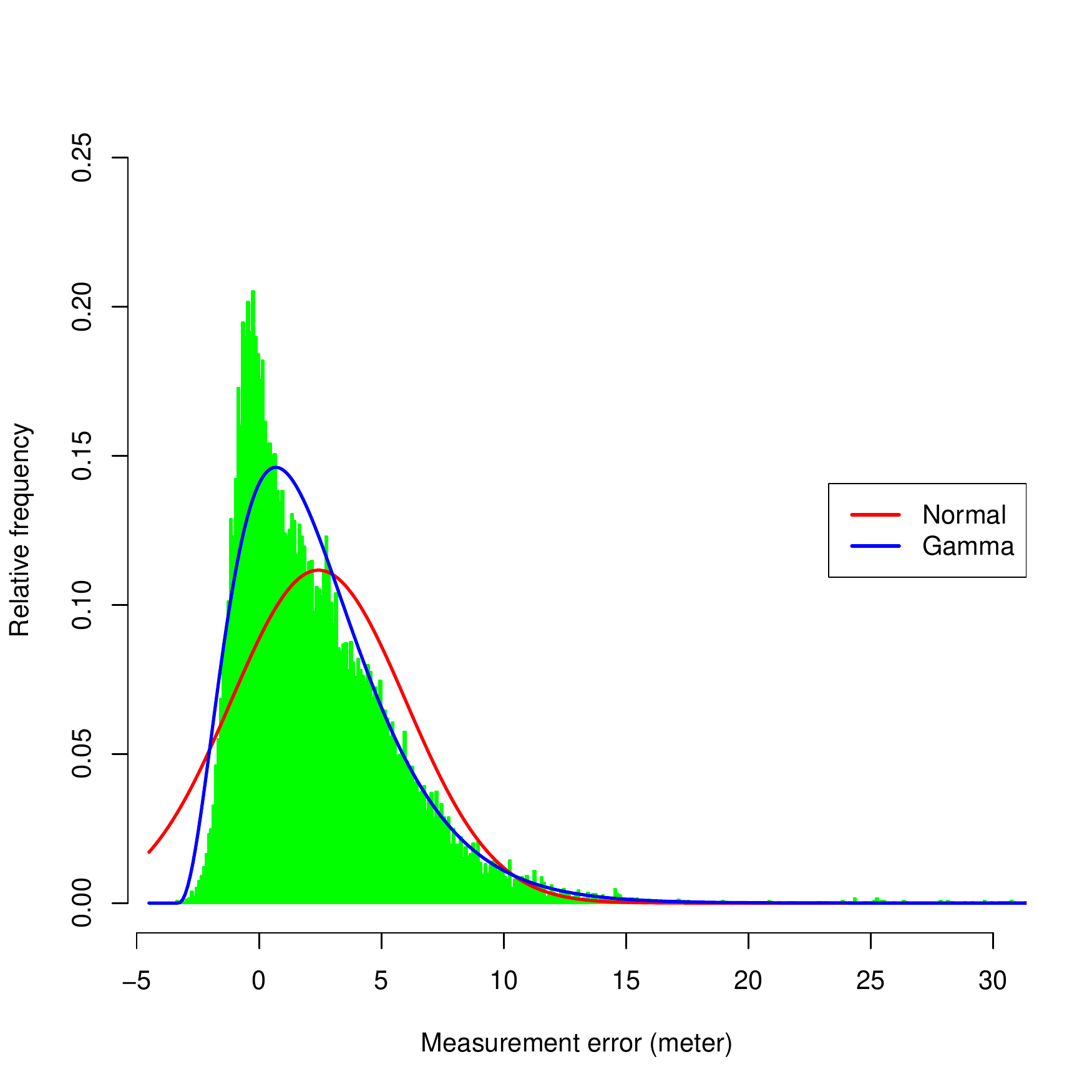}
  \caption[Normalized histogram of the distance measurement error of experiment Mobile 2 with 22901 successful \ac{TOF} measurements. When the distance measurement is shorter than the true distance, the error is defined as a negative error; otherwise as positive error. The mean absolute error is 2.85 meter and the skewness is 3.37. For display purposes, measurement errors greater than 30 meters are not included in the histogram although measurement errors can be has large has 75~m in rare cases. The fitted gamma distribution has a location parameter $\mu$ of -3.31~m.]{}
  \label{fig:histogram_ranging_and_fitting}
\end{figure}

We approximate the error distribution for real environment using normal distribution $\mathcal{N}(\mu,\sigma^2)$ and  gamma distribution $\Gamma(\alpha,\beta)$. Figure~\ref{fig:histogram_ranging_and_fitting} depicts the frequency histogram of the distance measurement error collected during an experiment with over 22000 \ac{TOF} measurements, having a positive biased and right-side tailed error. Figure~\ref{fig:histogram_ranging_and_fitting} also shows the distribution fitting of the error using the normal distribution $\mathcal{N}(2.43, 3.57^2)$ (red curve) and the gamma distribution $\Gamma(3.3, 0.58)$ (blue curve).

\subsubsection{Normal distribution}

Assuming a normal distribution with mean $\mu$ and variance $\sigma^2$, the likelihood function is given by \citep{Gezici:2008:SWP:1341571.1341575}:
\vspace{12pt}
\begin{equation}
p(r\mid u) = \frac{1}{(2\pi)^{N_{anc}/2}\left| C \right|^{1/2}}\exp\Bigl(-\frac{1}{2}(r - f(u)  - \mu)^T C^{-1}(r - f(u) - \mu)\Bigr)
\label{likelihood_function_gaussian}\vspace{12pt}
\end{equation}
where $C=\operatorname{diag}(\sigma^2_1, \sigma^2_2,\dots,\sigma^2_{N_{anc}})$ is the covariance matrix for $r$ in case of uncorrelated noise components and $f(u)=\|u - a\|$ is the noise-free distance vector. The \ac{ML} estimate is obtained by maximizing Eq.~\eqref{likelihood_function_gaussian}. For computational convenience, however, the \ac{ML} estimate is obtained by maximizing the log-likelihood function
\vspace{12pt}
\begin{equation}
\ln p(r\mid u) = \ln\Biggl(\frac{1}{(2\pi)^{N_{anc}/2}\left| C \right|^{1/2}}\Biggr) - \frac{1}{2}(r - f(u) - \mu)^T C^{-1}(r - f(u) - \mu)\text{.}
\label{likelihood_function_gaussian_ln}
\end{equation}

As the first term is independent of $u$, maximizing Eq.~\eqref{likelihood_function_gaussian_ln} is equivalent to minimizing the second term. The \ac{ML} estimate therefore is:
\vspace{12pt}
\begin{equation}
\begin{split}
\hat{u}_{\textit{ML}} &= \underset{u}{\operatorname{argmin}} \text{ } (r - f(u) - \mu)^T C^{-1}(r - f(u) - \mu)\\
&=\underset{u}{\operatorname{argmin}} \text{ }\sum_{i=1}^{N_{anc}} \frac{\Bigl(r_i - \|u - a_i\| - \mu_i\Bigr)^2}{\sigma_i^2}\text{.}
\end{split}
\label{likelihood_function_gaussian_ln_estimate}
\end{equation}

Common techniques for solving Eq.~\eqref{likelihood_function_gaussian_ln_estimate} include numerical methods like Newton--Raphson procedure, Gauss--Newton method or steepest descent algorithm \citep{zekavat2011handbook}. For a noise distribution with zero mean the minimization of Eq.~\eqref{likelihood_function_gaussian_ln_estimate} corresponds to a weighted version of the NLLS algorithm where the weights are inversely proportional to the noise variances thus larger variances result in smaller weights. In general, the \ac{ML} estimator generalizes the NLLS method and is reduced to it when assuming zero mean and when all $\sigma_i^2$ are identical. For the rest of this paper the algorithm is referred to as MLE-$\mathcal{N}$.

\subsubsection{Gamma distribution}

Looking at the histogram in Figure~\ref{fig:histogram_ranging_and_fitting}, a gamma distribution looks like a better choice for calculating the likelihood. Like the histogram of distance measurements, the gamma distribution is asymmetric, has only positive support, and still allows arbitrary large distance measurement errors with decreasing probability.

The gamma distribution is parametrised by a shape parameter $\alpha\geq 0$ and a rate parameter $\beta\geq 0$. The \ac{pdf} is given by:
\vspace{12pt}
\begin{equation}
  \label{eq:crlb:pdf:gamma}
  \Gamma(\alpha,\beta)(x)=\frac{\beta^\alpha}{\Gamma(\alpha)}x^{\alpha-1}e^{-\beta x}
  \vspace{12pt}
\end{equation}
where $\Gamma(\alpha)=\int_{0}^{\infty}t^{\alpha-1}e^{-t}dt$.

Since the gamma distribution is not defined for non-positive values and the distance measurements are sometimes too short, we add an offset $\eta\geq 0$ to the measurements, which is chosen to be the smallest observed measurement error. The likelihood to be at position $u$ while measuring $r_i$ from anchor $a_i$ is:
\vspace{12pt}
\begin{equation}
  p_i(r_i\mid u)=
  \begin{cases}
    \Gamma(\alpha,\beta)(r_i+\eta-\|u-a_i\|)&\text{if $r_i\geq\|u-a_i\|-\eta$}\\
    0 & \text{otherwise}
  \end{cases}\ .
  \vspace{12pt}
\end{equation}
Approximating the maximum joint likelihood $p(r\mid u)=\prod_ip_i(r_i\mid u)$ for position $u$ with respect to measurements turns out to be tricky. If the position $u$ is too close to an anchor, the joint likelihood of $u$ and some environment of $u$ is $0$, as is the gradient $\nabla \prod_ip_i(r_i\mid u)$. Thus, using log-likelihood and the method of steepest descent will fail. Instead, an iterative method that does not rely on the gradient of $p(r\mid u)$ is more successful. In degenerate cases, the result will be the initial guess, while a maximum likelihood can usually be approximated. For the rest of this paper the algorithm will be denoted as MLE-$\Gamma$.

\newpage
\section{The Membership Degree Min-Max Algorithm}
\label{section:md-min-max}
 
Based on the previous work and our experiment results, Min-Max is a potential method for position estimation with the inexact range measurements.
Different from most algorithms estimating the position based on an exact mathematical derivation or probability, we propose \ac{MD-Min-Max} algorithm. \ac{MD-Min-Max} employs an empirical \acl{MF} (\acs{MF}) to convert range measurements into degrees of support on the four vertices ($\mathbf{V}=\{v_j\mid j\in{\downarrow 4}\}$) obtained by Min-Max.

For any partially ordered set $(P,\leq)$ and any $p\in P$ we define the downset ${\downarrow p}=\{q\in P\mid q\leq p\}$. From now on, we use the partially ordered set $(\mathbb{N}^+,\leq)$ of positive integers. For example, ${\downarrow 4}=\{1,2,3,4\}$.

\subsection{Concepts of Fuzzy Set}
 
\begin{figure}[htb]
\centering
\includegraphics[width=2.0in]{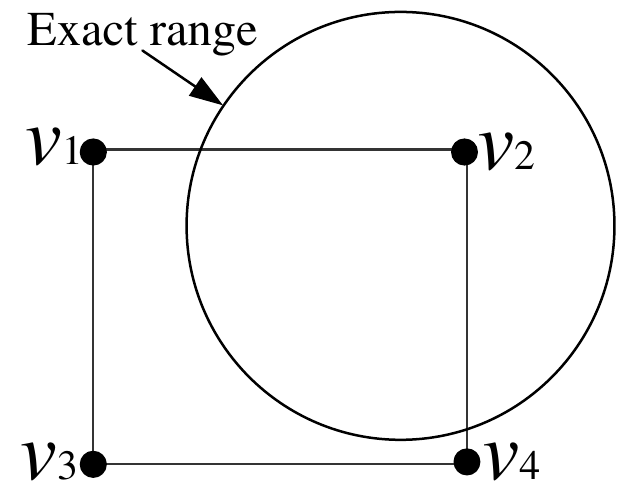}
\caption[The relationship between range and vertex.]{}
\label{fig:prob}
\end{figure}

Since range measurements $r$ of indoor scenarios are uncertain and imprecise, $r$ cannot estimate the target position determinately. Probability theory is the most common way to deal with uncertainty, however, it requires the \ac{pdf} of measurements and incurs high computation. More important, it cannot present the relationship between the estimated position $(\hat{x},\hat{y})$ and the defined set $\mathbf{V}$ of Min-Max. For example, if given an exact range $\bar{r}$ as Figure~\ref{fig:prob}, the weighting function $W_a(j)$ of \ac{E-Min-Max} is able to present the difference between $(\hat{x},\hat{y})$ and $\mathbf{V}$ in ideal case, however, it treats all residues equally under uncertain range errors. Probability method fails in ideal case, because $\bar{r}$ is a determinate event rather than a random variable. Fuzzy set \citep{zadeh1965fuzzy} (like 'nearby' or 'distant' of positioning) and evidences (like all range measurements) are able to describe the nearness between $(\hat{x},\hat{y})$ and $\mathbf{V}$. In Figure~\ref{fig:prob}, the result of using fuzzy set is that $(\hat{x},\hat{y})$ is nearby $v_1$ and $v_4$ but far from $v_2$ and $v_3$. Overall, fuzzy concept is more suitable to describe the relationship of $(\hat{x},\hat{y})$ to $\mathbf{V}$ obtained by Min-Max.

\subsection{Membership degree}

Of fuzzy set, we use the concept of \emph{membership degree} \citep{zadeh1965fuzzy} ($\mu(\bar{d})$) only. Here, membership degree means that one fuzzy variable partially belongs to a fuzzy set, then the estimated position is close to the $\mathbf{V}$. In \ac{MD-Min-Max} algorithm, the normal localisation formulas are replaced by rules. To make the algorithm simple and fast, we only employ one $\mathbf{rule}$ in the convert step in Algorithm~\ref{alg:md-min-max}, which presents the agreement of $(\hat{x},\hat{y})$ belonging to $\mathbf{V}$:

\begin{equation}
\label{eq:fuzzy-rule}
\begin{array}{l}
\mathbf{If}\ \left\| {v_j-\mathrm{a}} \right\| \text{ approximates } r,\\
\mathbf{then}\ (\hat{x},\hat{y})\ \text{ is nearby }v_j.\\
\end{array}
\end{equation}

A numerical value in the interval $[0,1]$ stands for the degree of agreement in Eq.\ \eqref{eq:fuzzy-rule}, and is calculated by \ac{MF}. The higher the degree is, the higher is the agreement in Eq.\ \eqref{eq:fuzzy-rule}. To show the intuition, consider Figure~\ref{fig:prob}. Eq.~\eqref{eq:fuzzy-rule} should result in a higher membership degree of $v_1$ and $v_4$ than for $v_2$ and $v_3$.
\vspace{12pt}
\begin{example}
  \begin{gather*}
    \mathbf{If}\ r-\left\| {v-\mathrm{a}} \right\| \text{ is } 1.0 \ \mathbf{then}\ r \text{ supports } v \text{ by a degree of } 0.6 \\
    \mathbf{If}\ r-\left\| {v-\mathrm{a}} \right\| \text{ is } 0.0 \ \mathbf{then}\ r \text{ supports } v \text{ by a degree of } 1.0 \\
    \mathbf{If}\ r-\left\| {v-\mathrm{a}} \right\| \text{ is } -0.3 \ \mathbf{then}\ r \text{ supports } v \text{ by a degree of } 0.9
  \end{gather*}
\end{example}
 
\begin{figure}[htb]
\centering
\includegraphics[width=.70\linewidth]{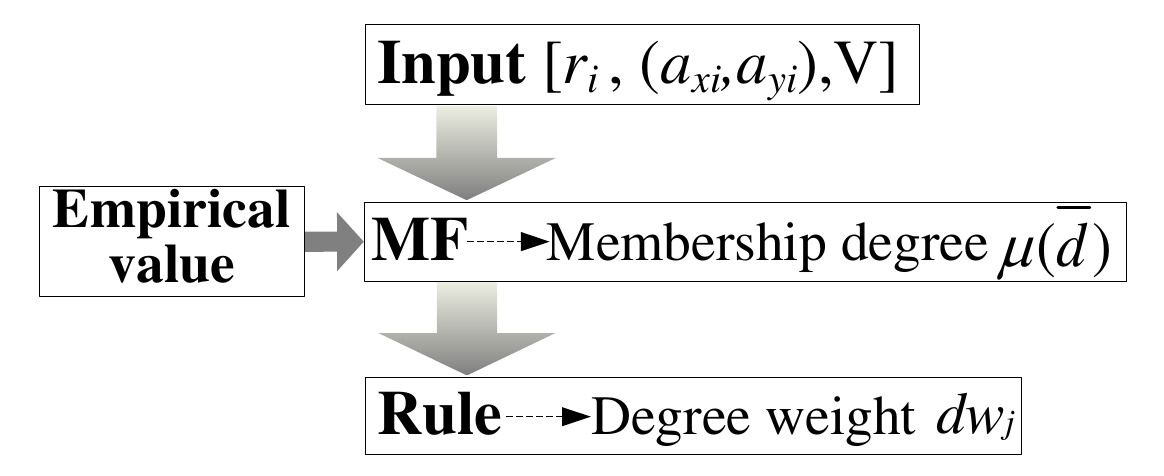}
\caption[The framework of  membership degree in \ac{MD-Min-Max}, with $i \in {\downarrow N_{\mathrm{anc}}}$ and $j \in {\downarrow 4}$.]{}
\label{fig:fuzzprocedure}
\end{figure}

The framework to involve the membership degree on Min-Max is shown in Figure~\ref{fig:fuzzprocedure}, and the procedure of \ac{MD-Min-Max} is in Algorithm~\ref{alg:md-min-max}.

\begin{algorithm}
  \caption{\ac{MD-Min-Max}}\label{alg:md-min-max}
  \begin{algorithmic}[1]
    \Require{ranges $r_i$ and anchor positions $a_i$, $i \in {\downarrow N_{\mathrm{anc}}}$, and vertices $\{v_j\mid j\in{\downarrow 4}\}$ computed by Min-Max;}
    \Ensure{the estimated position $(\hat{x},\hat{y})$}
    \For{$i\in{\downarrow N_{\mathrm{anc}}}$}\Comment{Compute membership degrees}
    \For{$j\in{\downarrow 4}$}
    \State Compute $d_{ij}=|| {v_j-a_i}||$
    \State Compute $\bar{d}_{ij}$ by Eq.\ \eqref{eq:absoluted};
    \State Calculate membership degree $\mu(\bar{d}_{ij})$ by Eq. \eqref{eq:mu};
    \EndFor
    \EndFor
    \For{$j\in{\downarrow 4}$}    
    \State Calculate degree weight $\mathit{dw}_j$ by Eq.\ \eqref{eq:rulemath};
    \EndFor
    \State Estimate position $(\hat{x},\hat{y})$ as weighted average by Eq.\ \eqref{eq:aggregated1};
  \end{algorithmic}
\end{algorithm}

\subsection{Membership function}
\label{membership_function}

Typically, membership functions are defined by experts or generated from statistics. We suppose that the error distribution of distance measurements in the same scenario are similar, thus the \ac{MF} can be configured by empirical values obtained from previous experiments in the same scenario. The empirical knowledge involved in \ac{MF} helps in making the algorithm adaptive to conditions of imprecise distance measurements. The triangular \ac{MF} is determined by three parameters ($\mathit{MF}_{\mathrm{low}}, \mathit{MF}_{\mathrm{median}}, \mathit{MF}_{\mathrm{up}}$), where (($\mathit{MF}_{\mathrm{low}},0$), ($\mathit{MF}_{\mathrm{median}},1$), ($\mathit{MF}_{\mathrm{up}},0)$) are the three vertices of the triangular \ac{MF}. We calculate the three parameters of \ac{MF} as follows:
\begin{enumerate}
\item Obtain a large number of samples of range measurements $r$ and the corresponding reference ranges $\overline{r}$;
\item Compute the median value of all $r-\bar{r}$, named as $\mathit{MF}_{\mathrm{median}}$;
\item Compute $\mathit{MF}_{\mathrm{up}}$ as the $0.995$ quantile and $\mathit{MF}_{\mathrm{low}}$ as the $0.005$ quantile;
\item Configure the triangular \ac{MF} with three parameters $((\mathit{MF}_{\mathrm{low}},0)$, $(\mathit{MF}_{\mathrm{median}},1)$, $(\mathit{MF}_{\mathrm{up}},0))$.
\end{enumerate}

\subsubsection{Analysing range measurements}
\label{range_measurements}

\begin{figure}[htb]
  \subfloat[Number of successful measurements for distance]{\label{fig:hist-distances}\includegraphics[width=.50\linewidth]{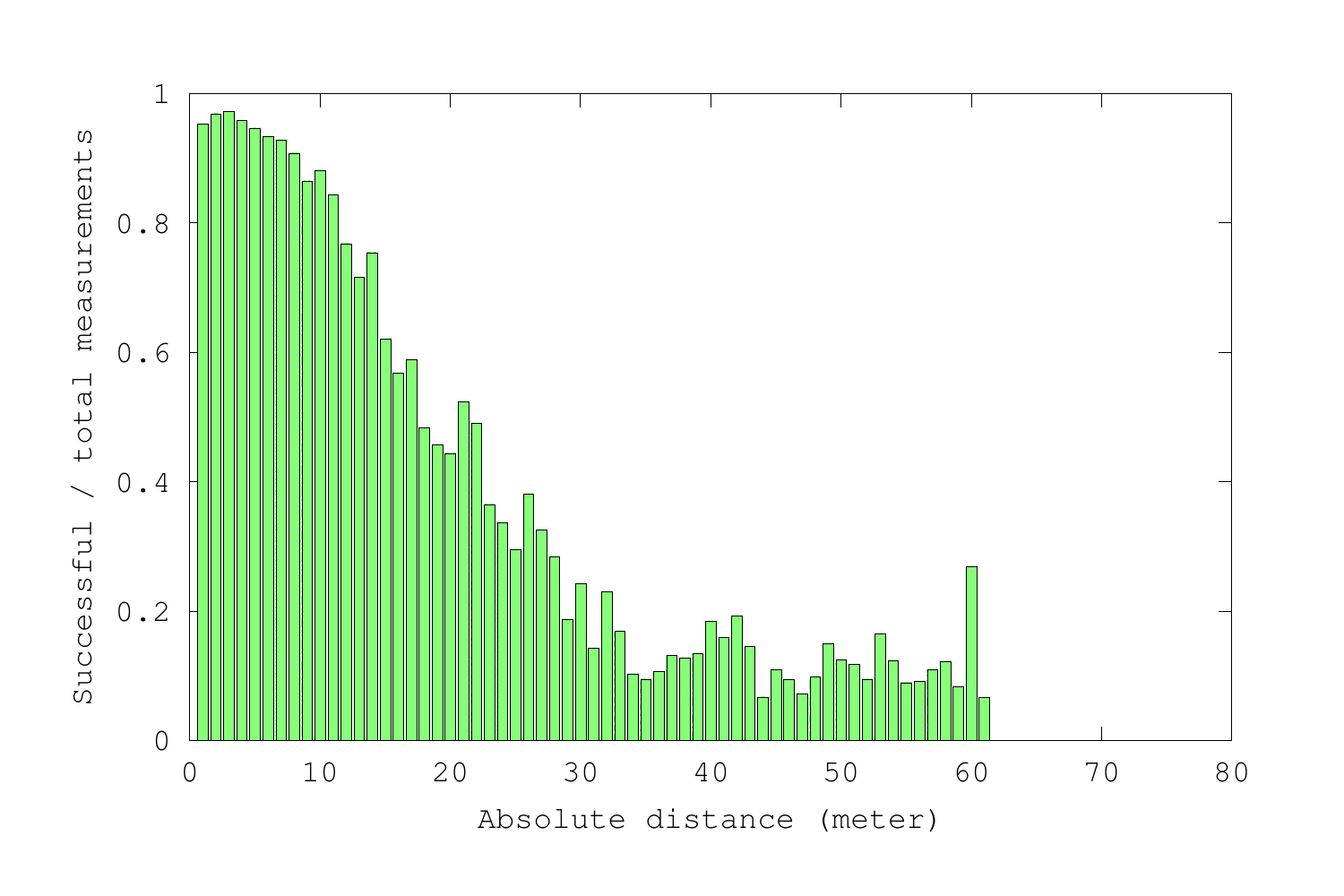}}
  \subfloat[Absolute range errors for distance]{\label{fig:range-abs-error}\includegraphics[width=.50\linewidth]{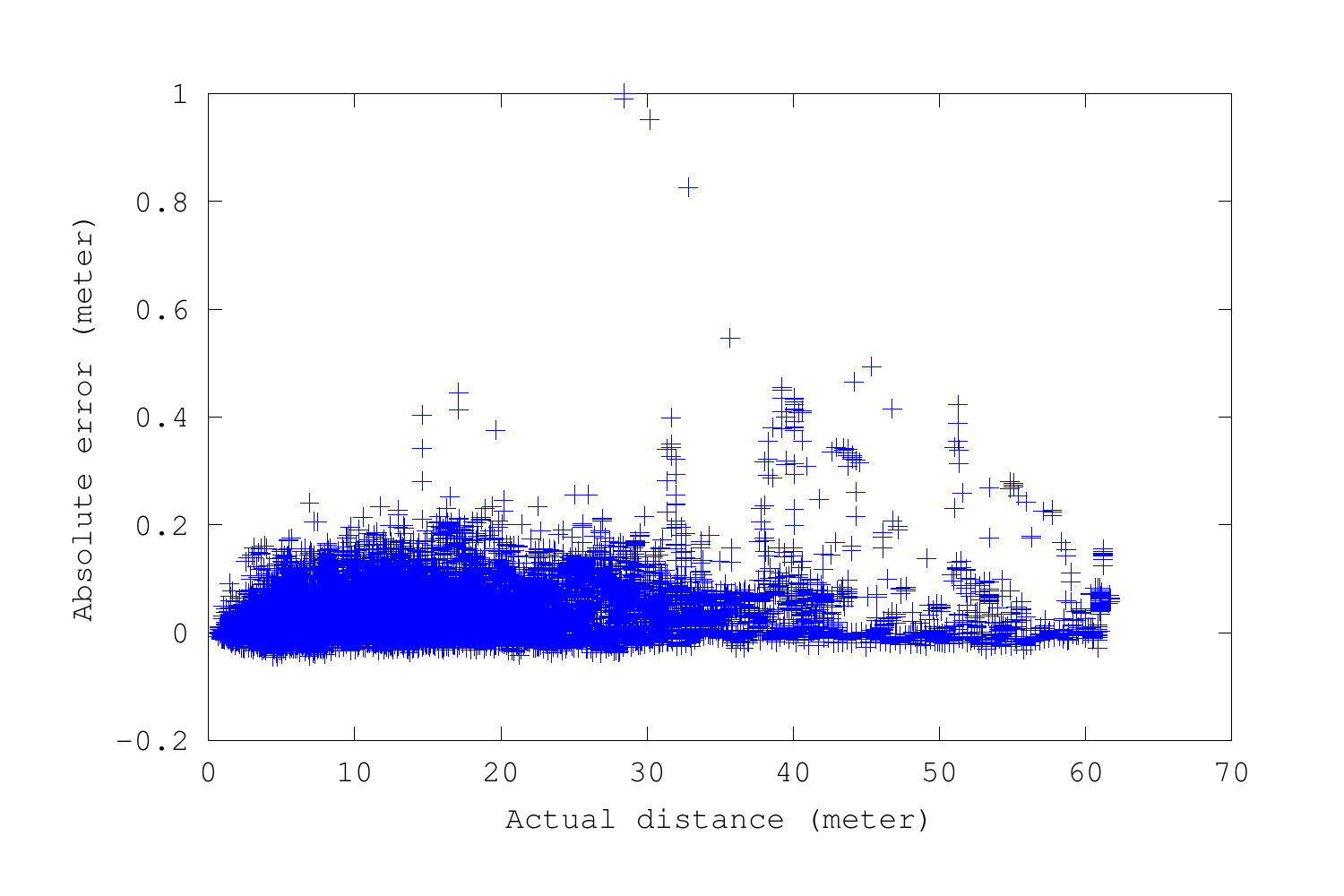}}\\
  \caption[Statistical modelling]{}
  \label{fig:statistical-modelling}
\end{figure}

The first step of computing a \ac{MF} is to obtain a large number of range measurements using a reference system. For this example, we conducted an experiment where we used a robot to provide us with reference locations and collected range measurements. The experiment involved $17$ anchors placed into an office building. Each anchor was ranged $3043$ times. Since some measurements failed, we collected $22901$ distance measurements at varying distances from the anchor nodes. Figure~\ref{fig:hist-distances} displays the relative number of successful measurements at those distances. Figure~\ref{fig:range-abs-error} displays the distribution of absolute measurement errors. As one can see, the distance error is independent of the distance we measured. At short distances, measurements occur to be more noisy, with outliers at medium distances. Outlier values have been measured at a distance of about $27$ meters and $37$ meters. The lack of many large range errors at more than about $37$ meters is explained by the high probability of a failed measurement. Distance measurements of $30$ meters and longer succeeded in only $20\%$ to $30\%$ of the attempts. Figure~\ref{fig:range-abs-error} shows the distribution of the absolute errors for the measured distances. As we see, the absolute error is uncorrelated to the actual distance. Indeed, the correlation is $0.1373$. This is a strong evidence that both quantities are independent and our \ac{MF} can be formulated independent of the range, i.e.\ in terms of absolute errors.

\subsubsection{Configure \ac{MF}}

\begin{figure}[htb]
  \centering
  \includegraphics[width=0.70\linewidth]{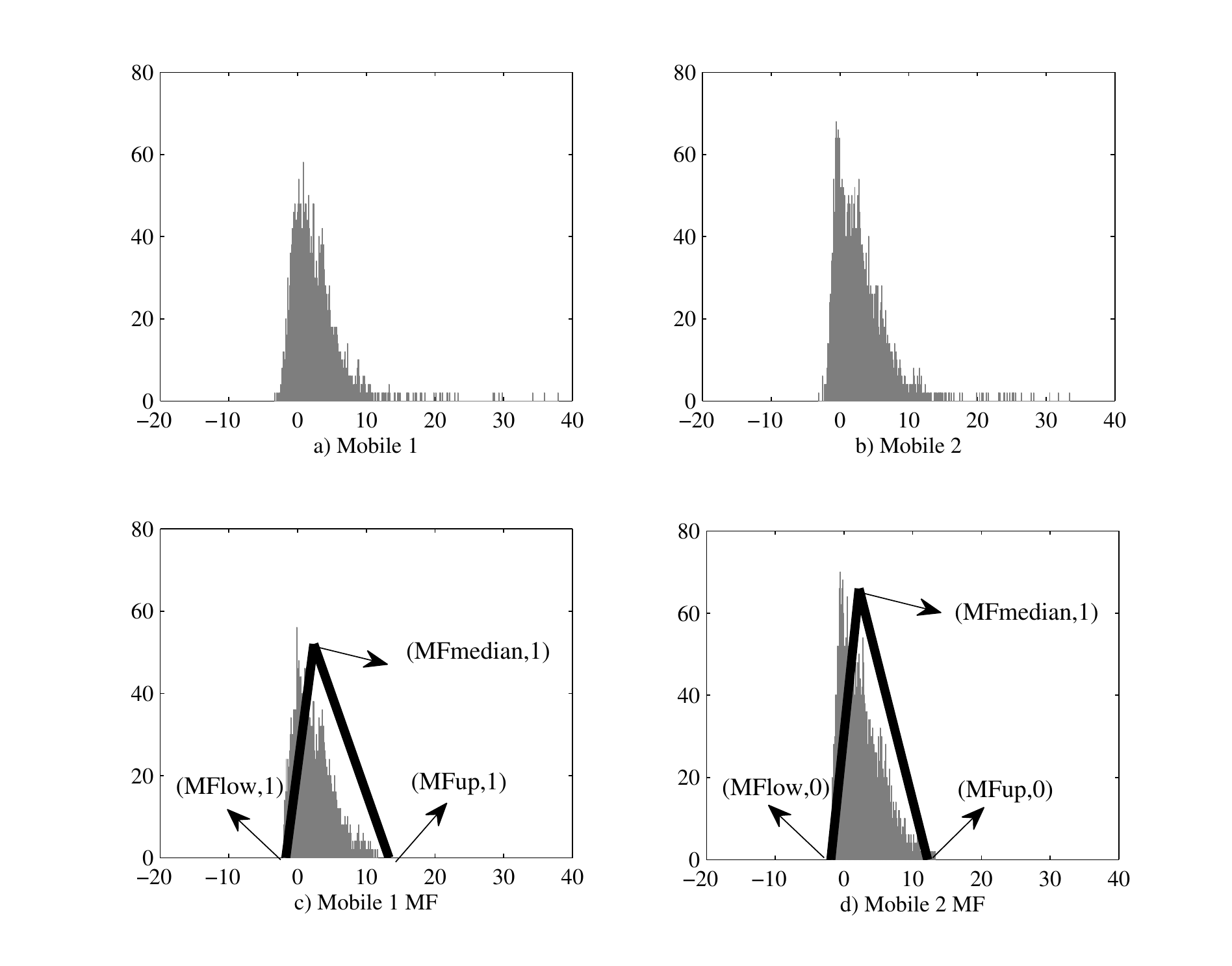}
  \caption[Frequency histogram of absolute range ($\mathrm{Fr}(r-\overline{r})$) is collected from two experiments (Mobile 1 and Mobile 2), where $\overline{r}$ is the set of value of reference range and $r$ is the measured range. Figure~\ref{fig:Statistics} (a-b) illustrate the histogram of $\mathrm{Fr}(r-\overline{r})$, and Figure~\ref{fig:Statistics} (c-d) is the histogram after discarding the outliers. The triangle profiles in Figure~\ref{fig:Statistics} (c-d) are the \ac{MF} determined by three parameters ($\mathit{MF}_{\mathrm{low}}, \mathit{MF}_{\mathrm{median}}, \mathit{MF}_{\mathrm{up}}$).]{}
  \label{fig:Statistics}
\end{figure}

We performed two experiments, named as Mobile 1 and Mobile 2, with a mobile node moving along two different routes in the same office building. Then, a absolute range error histogram of the measurements is used to configure the \ac{MF}. The histogram presents range measurements in absolute form ($r-\bar{r}$), where $\bar{r}$ is the distance obtained by our reference system and $r$ is the measured distance. Figure~\ref{fig:Statistics} (a-b) shows the histograms for our experiments. Here, we use a triangular \ac{MF}. Its empirical parameters are shown in Figure~\ref{fig:Statistics} (c-d). 

The \ac{MF} parameters of Mobile 1 and Mobile 2 are $[-1.7, 2.38, 13.31]$ and $[-2.161, 1.636, 16.043]$ separately, which also indicates that distance in the same scenario maintains familiar behaviour. Thus, configuring the \ac{MF} based on empirical values is reasonable, making this algorithm easy to implement in other indoor scenarios.

The triangular \ac{MF} is not the only type of \ac{MF} fitting our \ac{MD-Min-Max} algorithm. Also other \ac{MF}, such as trapezoidal \ac{MF}, quadratic function \ac{MF}, rectangular \ac{MF}, or any other theoretical distribution of the statistical result can be used. \ac{MD-Min-Max} employs triangular \ac{MF} because of its conceptual simplicity and computational efficiency. For different scenarios, the \ac{MF} should be chosen according to its range condition, as approximate to the frequency histogram as possible.

\subsection{Convert Range into Membership Degree by \ac{MF}}

\ac{MD-Min-Max} first computes the four vertex coordinates $\mathbf{V}$ of \ac{IR} by Min-Max. Given the coordinates of the anchors, it is simple to compute the distance between the $i$th anchor and $j$th vertex: $d_{ij}=\left\| {v_j-a_i} \right\|$. Since the \ac{MF} is expressed in absolute measurement errors, the distance between a vertex and an anchor is also described as an absolute difference between measurement $r_i$ and $d_{ij}$, as shown in Eq.~\eqref{eq:absoluted}.
\vspace{12pt}
\begin{equation}\label{eq:absoluted}
  \bar{d}_{ij}=r_i-d_{ij}=r_i-||{v_j-a_i}||
        \text{ for $i \in {\downarrow N_{anc}}$, $j \in {\downarrow 4}$.}
\end{equation}

Then, the \ac{MF} $\mu(\bar{d})$ can be used to calculate an agreement degree $\mu(\bar{d}_{ij})$, as shown in Eq.~\eqref{eq:mu}. The range of membership degree is a real number between zero and one. It is characterized by three parameters $[\mathit{MF}_{\mathrm{low}}, \mathit{MF}_{\mathrm{median}}, \mathit{MF}_{\mathrm{up}}]$ which are obtained from the previous empirical data. For example, for positioning in Mobile~1 case, we should use the parameters obtained by the samples of Mobile~2.

\begin{equation}\label{eq:mu}
  \mu(\bar{d})=
  \begin{cases}
    \frac{\bar{d}-\mathit{MF}_{\mathrm{up}}}{\mathit{MF}_{\mathrm{median}}-\mathit{MF}_{\mathrm{up}}}&
    \text{if $\mathit{MF}_{\mathrm{median}} \leq \bar{d} < \mathit{MF}_{\mathrm{up}}$} \\
    \frac{\bar{d}-\mathit{MF}_{\mathrm{low}}}{\mathit{MF}_{\mathrm{median}}-\mathit{MF}_{\mathrm{low}}}&
    \text{if $\mathit{MF}_{\mathrm{low}} < \bar{d} < \mathit{MF}_{\mathrm{median}}$} \\
    0 & \textrm{otherwise} \\
  \end{cases}
\end{equation}

Eq.~\eqref{eq:mu} describes that the membership degree $\mu_{ij}=\mu(\bar{d}_{ij})$ decreases from $1$ to $0$ as $\bar{d}_{ij}$ moves away from $\mathit{MF}_{\mathrm{median}}$; to be more specific, if $\bar{d}_{ij}$ is outside of the interval $[\mathit{MF}_{\mathrm{low}},\mathit{MF}_{\mathrm{up}}]$, then $\mu_{ij}$ is $0$. If a range measurement $r_i$ is severely corrupted, then $\bar{d}_{ij}$ is very far from $\mathit{MF}_{\mathrm{median}}$ and $\mu(\bar{d}_{ij})$ is $0$. This is the case when the range measurement is considered to be an outlier.

A huge error between multiple ranges is uncommon as illustrated by the statistics in Figure~\ref{fig:Statistics} (a-b) and shown in several publications \citep{guvenc2007analysis,venkatraman2004novel,Venkatesh2006}. Overall, the greater the deviation from $\mathit{MF}_{\mathrm{median}}$, the higher the possibility that the range measurement has a large error. Therefore, the membership degree can averagely weaken these ranges as the long tail component in Figure~\ref{fig:Statistics} (a-b).

\subsection{Combine membership degree}

Since multiple ranges determine one estimation jointly, a conjunctive rule is made to combine multiple membership degrees into the weight on each vertex: $\mathit{dw}_j$, $j\in{\downarrow 4}$. The linguistic rule for the $j$th consequent is expressed as:
\vspace{12pt}
\begin{equation}
  \label{eq:rulelinguistic}
  \begin{array}{l}
   \textbf{If}\ \mu(\bar{d}_{ij}) \text{ fully agree to } v_j, i \in {\downarrow N_{\mathit{anc}}} \\
   \textbf{then}\ \mathit{dw}_j \text{ totally supports } v_j=(\hat{x},\hat{y}).\\
  \end{array}
\end{equation}

The signal-to-noise ratio in Eq.~\eqref{eq:rulemath}, defined as the reciprocal of the coefficient of variation of multiple degrees, is used as the weight of each vertex. The signal-to-noise ratio can be interpreted as a measure of the homogeneity of the range measurements and as the degree of agreement.
\vspace{12pt}
\begin{equation}
  \label{eq:rulemath}
  \mathit{dw}_j=
    \begin{cases}
      \frac{{\operatorname{mean} (\bigcup_{i = 1}^{N_{\mathit{anc}}} {\mu_{ij}})}}{\sqrt{\operatorname{var} (\bigcup_{i = 1}^{N_{\mathit{anc}}} {\mu_{ij}})}} & \text{if $\operatorname{var} (\bigcup_{i = 1}^{N_{\mathit{anc}}} {\mu_{ij}})> 0$} \\
      \infty & \text{if $\operatorname{var} (\bigcup_{i = 1}^{N_{\mathit{anc}}} {\mu_{ij}})= 0$}
    \end{cases}
\end{equation}

Thus, the larger $\mathit{dw}_j$ is (the higher the mean agreement and the smaller the agreement variance are), the more likely of target should be the vertex $v_j$. Combining multiple degrees in this way is not only simple, but also associates the conjunctive opinion of all ranges.

\subsection{Weighted average of~$\mathbf{V}$}

The final estimated position is the average of the four vertex coordinates weighted by their associated degree, as expressed in Eq.~\eqref{eq:aggregated1}.
\vspace{12pt}
\begin{equation}
  \label{eq:aggregated1}
  (\hat{x},\hat{y}) =\sum_{i=1}^{4} \frac{\mathit{dw}_i}{\sum_{j=1}^{4} \mathit{dw}_j} \cdot (v_{xi},v_{yi})
  \vspace{12pt}
\end{equation}
Vertices with higher accumulated degree and smaller degree variance are weighted higher. Therefore, the final estimation is considered to be a likely position within the four vertices of Min-Max, because of the good understanding of range errors derived from empirical knowledge.

\subsection{Complexity}

The run-time and memory requirements of the \ac{MD-Min-Max} algorithm are modest.
\vspace{12pt}
\begin{proposition}
  The run-time complexity of \ac{MD-Min-Max} is in $\Theta(N_{\mathit{anc}})$.
\end{proposition}

\begin{proof}
  The run-time of \ac{MD-Min-Max} is clearly dominated by the loop in Step 1. Calculating the distance and the membership degree can be performed in constant time. Weighting the degrees by the mean and the standard deviation can be performed in constant time, if a method like Welford's~\citep{welford62} is used during step 1. The loop body is executed four times for each anchor. Step 4 and 5 are again constant time.\vspace{12pt}
\end{proof}

\begin{proposition}
  The space complexity of \ac{MD-Min-Max} is in $\Theta(N_{\mathit{anc}})$.
\end{proposition}

\begin{proof}
  Most memory is required to store the two coordinates of the anchor nodes and range measurements, namely $3N_{\mathit{anc}}$ registers. Additional space is needed to store the indexing variables. The three parameters of the membership degree function, the corners of the Min-Max calculation and the weights of the four corners. 
\end{proof}

The asymptotic time and space complexity of \ac{MD-Min-Max} is equal to the one of the traditional Min-Max. Our benchmarks show that the \ac{MD-Min-Max} algorithm is about 50\% slower than the \ac{E-Min-Max} algorithms and about 9 times slower than the original Min-Max algorithm. As Min-Max is such an inexpensive algorithm, and the number of anchors $N_{\mathit{anc}}$ is low for most scenarios, limited by technical limitations of radio communication and the distance intervals, \ac{MD-Min-Max} is a viable algorithm for sensor networks, especially if we compare it to more complex algorithms like the NLLS method.

\newpage
\section{Distribution of the Spatial Position Error}
\label{section_spatial_evaluation}

The \emph{spatial distribution of the positioning error}, or shorter, the \emph{spatial error distribution}, measures the expected positioning error at each location in space. It characterises the performance of the algorithm for a specific anchor set-up and shows where an algorithm is expected to perform well and where it does not perform.

To evaluate the spatial distribution of the positioning error, we executed every algorithm on each position of a $1000\times 1000$ unit sized grid $1000$ times in our LS\textsuperscript{2} simulation engine \citep{ls2,will2012ls2}. LS\textsuperscript{2} calculates the position error for every discrete point in the simulated area using an error model and an algorithm selectable by the user. In the first scenario we chose a very basic anchor set-up with four anchors placed in the four corners of the playing field. The inter anchor distance is much higher than in most real world scenarios and shows the performance of the evaluated algorithms in borderline situations. The resulting image consists of up to three differently coloured areas. The grey area indicates a position error between 100\% and 500\% of the expected distance measurement error value; the darker the area, the higher is the error. The green area (if present) indicates a position error lower than the expected distance measurement error; the darker the area, the lower is the error. In the blue area the error 
is 
higher than 500\% of the distance error and is cropped to achieve a better image contrast. The anchors are represented by the small red squares.

The green area is very important for cooperative localisation strategies in WSNs, because the position error stays in a reasonable range as long as the node remains in the green area. Otherwise the position error tends to grow much faster than expected because for each step of the recursive cooperation strategy, the resulting position error is added to the average distance error. If the resulting position error is larger than the average ranging error, it grows very fast. 

For this simulation we chose a Gaussian distributed error for the general noise simulation and an exponential distributed error to simulate \ac{NLOS} situations. The expected value of the distance measurement error is 5\% of the playing field width, the standard deviation is 1.5\%. A \ac{NLOS} error occurs with a probability of 10\% and adds an exponential error with rate $2$. The membership function of the MD-Min-Max was set up like described in \ref{membership_function}. The inter-anchor distance is 15 times higher than the expected distance error.

\begin{figure}[htb]
\centering
\begin{minipage}{0.8\linewidth} 
\subfloat[Min-Max]{
\includegraphics[width=0.47\linewidth]{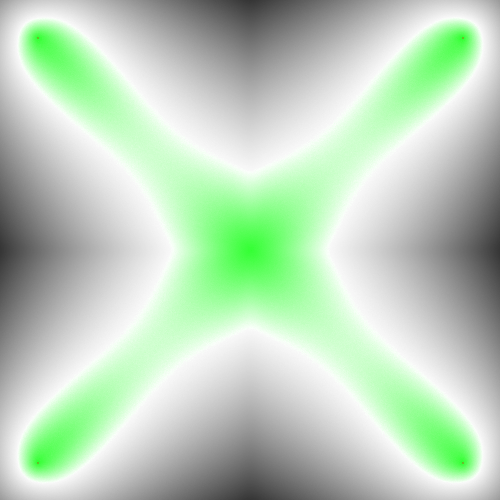}
\label{fig:subfig_4minmax}
}
\subfloat[\ac{E-Min-Max} (W4)]{
\includegraphics[width=0.47\linewidth]{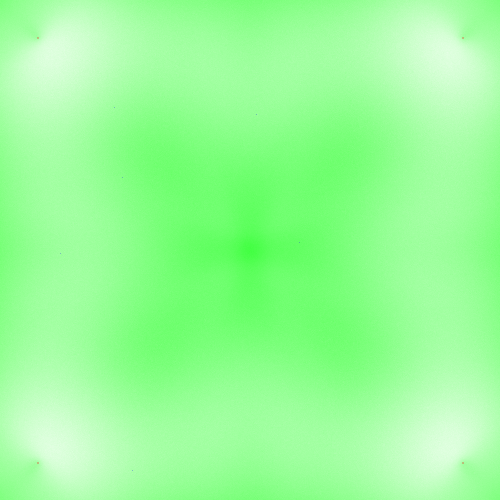}
\label{fig:subfig_4eminmaxw4}
} \\
\subfloat[\ac{E-Min-Max} (W2)]{
\includegraphics[width=0.47\linewidth]{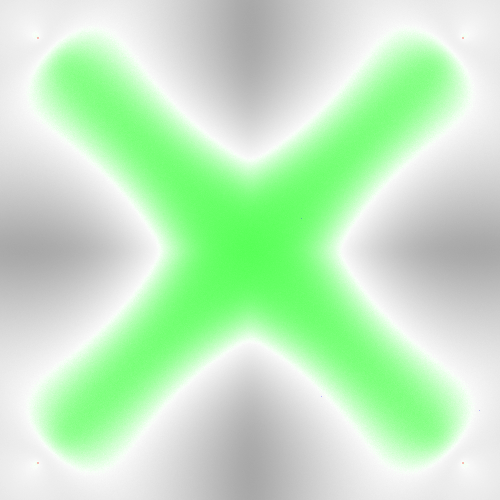}
\label{fig:subfig_4eminmaxw2}
}
\subfloat[\ac{MD-Min-Max}]{
\includegraphics[width=0.47\linewidth]{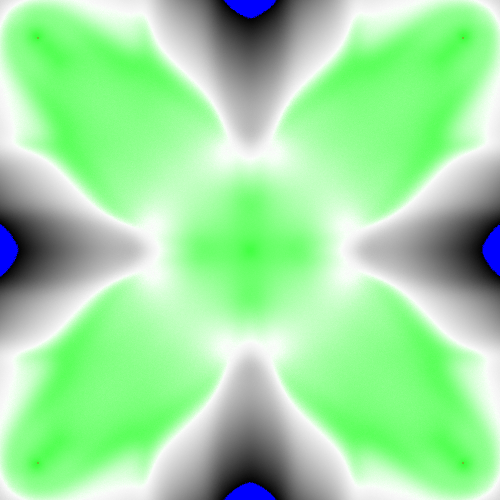}
\label{fig:subfig_4fgbloc}
}
\caption[Spatial distribution of the average position error with a basic anchor setup with one anchor in each of the four corners and a high inter-anchor distance.]{}
\label{fig:data4}
\end{minipage}
\end{figure}

We show the results of the first simulation run in Figure \ref{fig:data4}. The weaknesses of Min-Max are clearly visible. Min-Max performs very well only on the diagonal lines between the anchors and in the centre of the playing field. For similar setups in real world deployments Min-Max's performance is not really predictable because a mobile node will cross all areas. The \ac{E-Min-Max} (W2) algorithm performs slightly better in this setup but shows the same strengths and weaknesses. \ac{E-Min-Max} (W4) performs completely different in this scenario and shows a very homogeneous picture. It shows a slight performance drop close around the anchors but provides very good results for the rest of the area.  \ac{MD-Min-Max}'s results are comparable to Min-Max but with a slightly bigger area of high accuracy. Even if \ac{MD-Min-Max} has the highest accuracy inside the green area of all four algorithms one should choose \ac{E-Min-Max} (W4) for a random walk in such scenarios.

\begin{figure}[htb]
\centering
\begin{minipage}{0.8\linewidth} 
\subfloat[Min-Max]{
\includegraphics[width=0.47\linewidth]{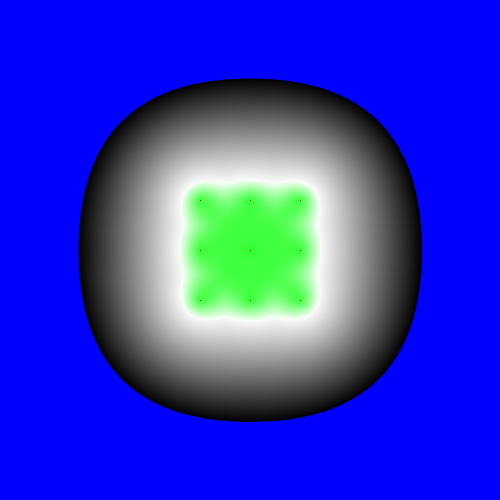}
\label{fig:subfig_1minmax}
}
\subfloat[\ac{E-Min-Max} (W4)]{
\includegraphics[width=0.47\linewidth]{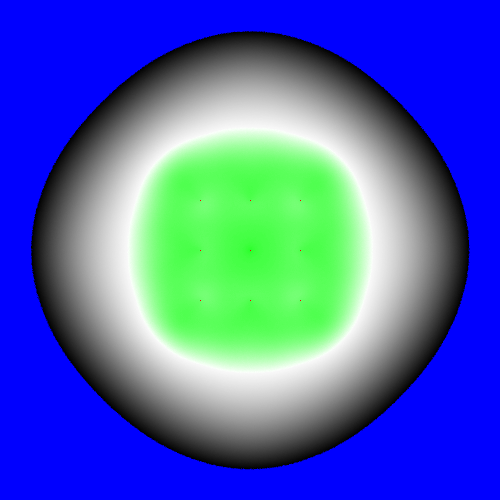}
\label{fig:subfig_1eminmaxw4}
} \\
\subfloat[\ac{E-Min-Max} (W2)]{
\includegraphics[width=0.47\linewidth]{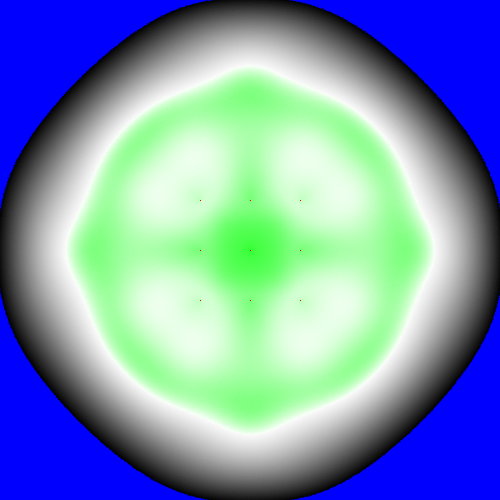}
\label{fig:subfig_1eminmaxw2}
}
\subfloat[\ac{MD-Min-Max}]{
\includegraphics[width=0.47\linewidth]{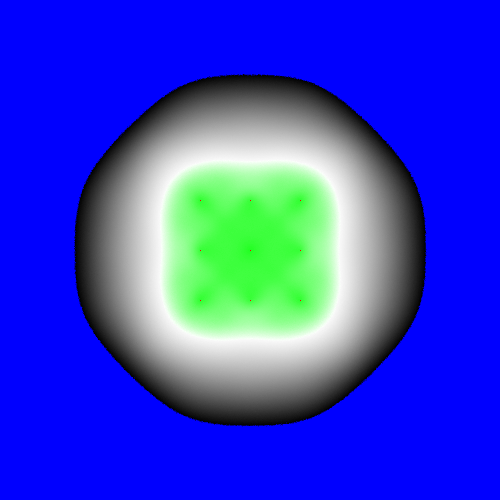}
\label{fig:subfig_1fgbloc}
}
\caption[Spatial distribution of the average position error with nine anchors concentrated in the middle of the simulation area and a very low inter-anchor distance.]{}
\label{fig:data1}
\end{minipage}
\end{figure}

The second simulation is shown in Figure \ref{fig:data1}. In this scenario we simulated every algorithm with a uniform grid layout for the anchors. We chose nine anchors which convex hull covers 4\% of the simulation area. The inter-anchor distance is comparable to common indoor deployments. The focus in this scenario is to evaluate how the algorithms will perform outside the convex hull of the anchors. The main strengths and the main weaknesses of Min-Max are clearly visible in this image. Min-Max performs very good inside the convex hull of a dense anchor setup and fast lowers its performance outside the convex hull down to unusable values. The main design goal of \ac{E-Min-Max} was to dilute this behaviour of Min-Max. As shown in Figure \ref{fig:subfig_1eminmaxw4} \ac{E-Min-Max} (W4) greatly improves the performance of Min-Max outside the convex hull without lowering the performance inside very much. \ac{E-Min-Max} (W2) stretches the usable area even a bit more but has some disadvantages in areas where Min-
Max performed well. Even if the average error over the whole playing field is nearly the same for both \ac{E-Min-Max} algorithms one could gain a noticeable advantage over the other if closer limitations to the area can be made in real world deployments. \ac{MD-Min-Max} clearly shows its advantages and disadvantages in this scenario. The area of high accuracy is only slightly increased and it also shows a fast performance drop outside the diagonals of the anchor hull, but the results inside this area are much more accurate than those of the Min-Max algorithm. For real world indoor deployments this observation can be important because the anchors are usually wall mounted and because of this, a mobile node rarely leaves the anchor hull.

\begin{figure}[htb]
\centering
\begin{minipage}{0.8\linewidth} 
\subfloat[Min-Max]{
\includegraphics[width=0.47\linewidth]{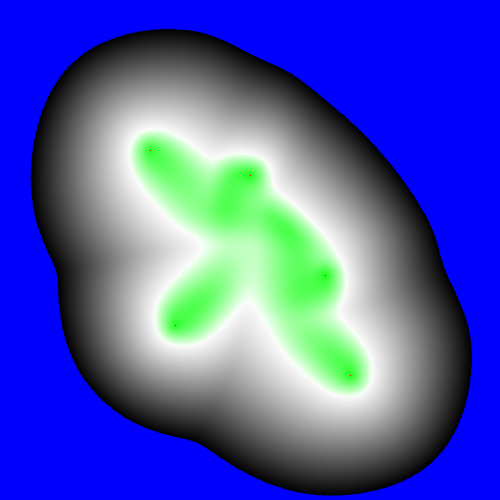}
\label{fig:subfig_3minmax}
}
\subfloat[\ac{E-Min-Max} (W4)]{
\includegraphics[width=0.47\linewidth]{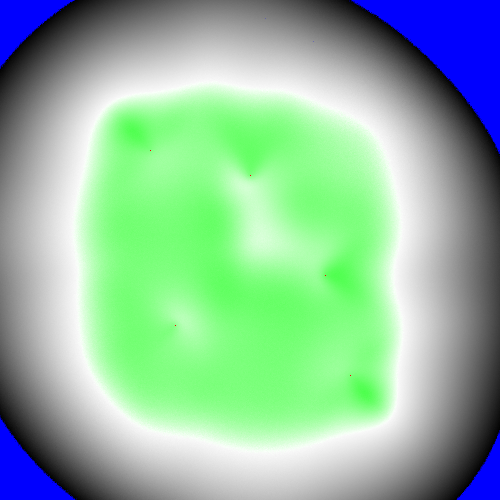}
\label{fig:subfig_3eminmaxw4}
} \\
\subfloat[\ac{E-Min-Max} (W2)]{
\includegraphics[width=0.47\linewidth]{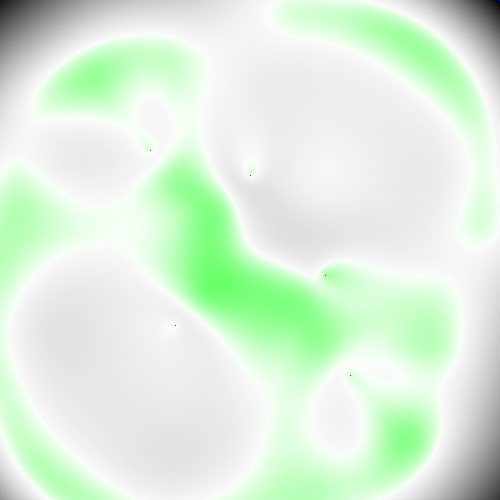}
\label{fig:subfig_3eminmaxw2}
}
\subfloat[\ac{MD-Min-Max}]{
\includegraphics[width=0.47\linewidth]{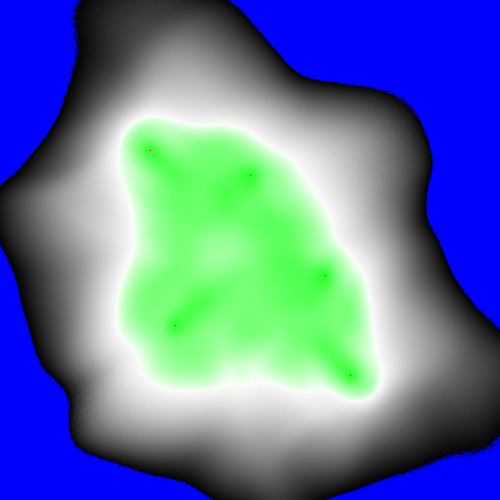}
\label{fig:subfig_3fgbloc}
}
\caption[Spatial distribution of the average position error with five anchor nodes in a more challenging setup with a medium inter-anchor distance.]{}
\label{fig:data3}
\end{minipage}
\end{figure}

In Figure \ref{fig:data3} the results of a more challenging scenario are shown. We placed four anchors nearly on a line and a fifth anchor to form a flat triangle with the rest. For most lateration algorithms this scenario is a kind of worst case scenario and the performance is weaker than the average performance of real world experiments because the overall number of anchors is low and the average inter-anchor distance is on a medium level. Min-Max has strong performance drops even inside the convex hull and then drops very fast to unusable values. \ac{E-Min-Max}~(W4) noticeably increases the performance and provides very good results for a centre area that covers ~30\% of the whole simulation area. \ac{E-Min-Max}~(W2) increases the average performance again but the results are very heterogeneous, so it could be challenging to make use of this performance gain in real world usage. \ac{MD-Min-Max} shows a comparable but much smaller shape than \ac{E-Min-Max}~(W4) but the accuracy inside this shape is much 
higher.

\begin{figure}[htb]
\centering
\subfloat[\ac{E-Min-Max} (W2) vs.\ Min-Max]{
\includegraphics[width=0.35\linewidth]{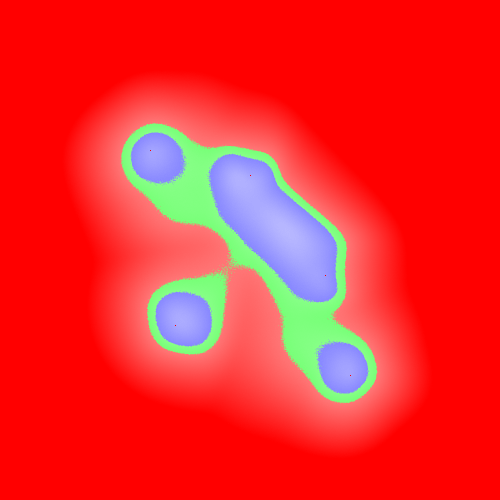}
\label{fig:diff-ew2-mm}
}\quad
\subfloat[\ac{E-Min-Max} (W4) vs.\ Min-Max]{
\includegraphics[width=0.35\linewidth]{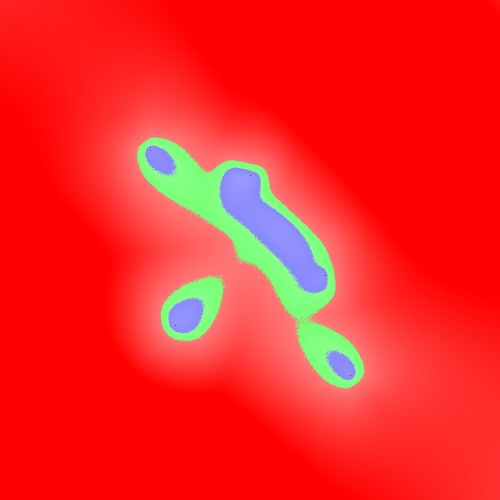}
\label{fig:diff-ew4-mm}
}\\
\subfloat[\ac{MD-Min-Max} vs.\ Min-Max]{
\includegraphics[width=0.35\linewidth]{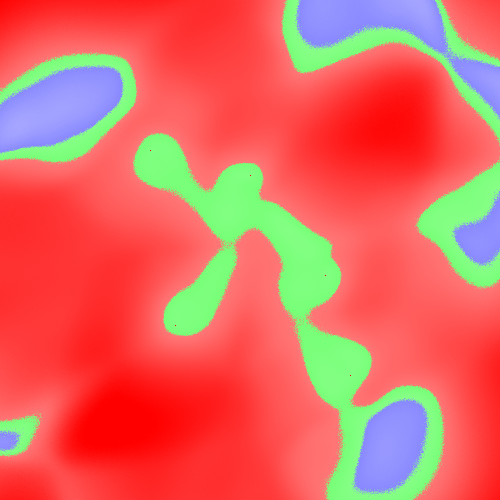}
\label{fig:diff-abs-mm}
}\quad
\subfloat[\ac{E-Min-Max} (W2) vs.\ \ac{E-Min-Max} (W4)]{
\includegraphics[width=0.35\linewidth]{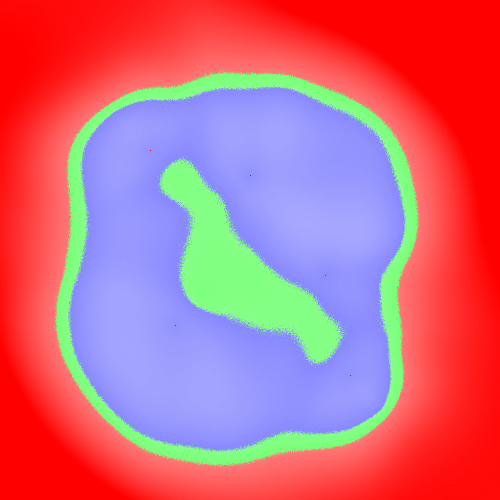}
\label{fig:diff-ew2-ew4}
}\\
\subfloat[\ac{MD-Min-Max} vs.\ \ac{E-Min-Max} (W2)]{
\includegraphics[width=0.35\linewidth]{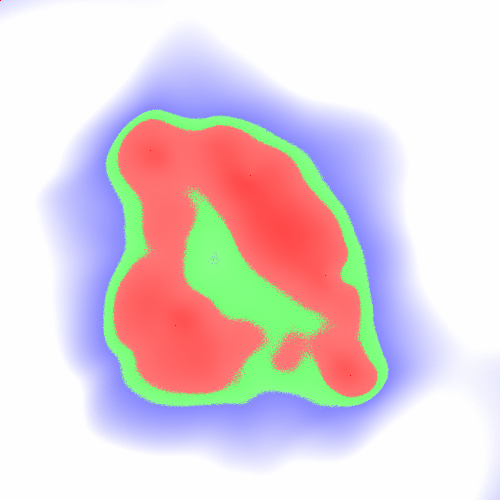}
\label{fig:diff-abs-ew2}
}\
\subfloat[\ac{MD-Min-Max} vs.\ \ac{E-Min-Max} (W4)]{
\includegraphics[width=0.35\linewidth]{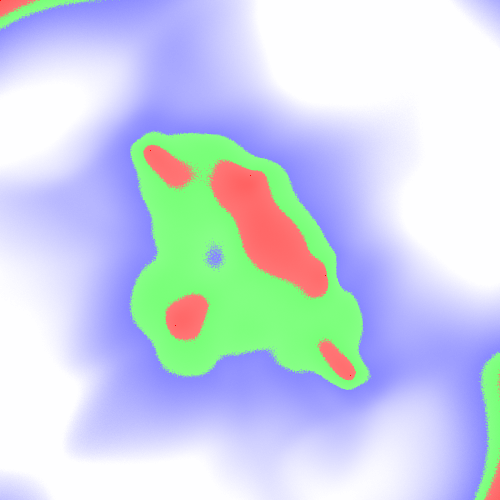}
\label{fig:diff-abs-ew4}
}
\caption[Comparing position errors of pairs of algorithms. Red areas indicate that the first algorithm outperforms the second, blue and white areas indicate that the second algorithm outperforms the first. Green areas indicate similar performance.]{}
\label{fig:diff}
\end{figure}

To highlight the difference of the average performance shown in Figure~\ref{fig:data3} between those algorithms, we visualize the difference of average errors between two algorithms in Figure~\ref{fig:diff}. Areas coloured in shades of red are areas in which the first mentioned algorithm achieves a lower average position error than the second algorithm. Areas coloured in shades of blue to white indicate areas in which the second algorithm achieves a lower position error. Areas coloured in green mark the areas in which both algorithms perform within $1.6\%$ of the playing field, i.e.\ their position error can be considered to be equivalent.

Figures~\ref{fig:diff-ew2-mm},~\ref{fig:diff-ew4-mm} and~\ref{fig:diff-abs-mm} show that the \ac{E-Min-Max} algorithms and our \ac{MD-Min-Max} algorithm all improve on Min-Max, especially outside of the area in which Min-Max performs best. The \ac{MD-Min-Max} algorithm is able to maintain the good performance of Min-Max in its strongest area and shows its weaknesses in areas outside of the convex hull of the anchors. Figure~\ref{fig:diff-ew2-ew4} compares \ac{E-Min-Max}~(W2) to \ac{E-Min-Max}~(W4) and shows that both can complement each other well. In the inner, blue tinted area, \ac{E-Min-Max}~(W4) compares much better while outside of that area, \ac{E-Min-Max}~(W2) performs better. Interestingly, their performance is comparable in the convex hull of the anchors, and thus worse than the original Min-Max algorithm. Figures~\ref{fig:diff-abs-ew2} and~\ref{fig:diff-abs-ew4} compare our \ac{MD-Min-Max} algorithm to \ac{E-Min-Max}~(W2) and (W4). Outside of the convex hull of the anchors, the \ac{E-Min-Max} algorithms perform much better than \ac{MD-Min-Max} however, in the centre area, performance is comparable or \ac{MD-Min-Max} is able to reduce the position error significantly. These areas, however, are of interest in many indoor deployments, where the mobile node is usually inside of the hull of anchors.

\begin{figure}[htb]
\centering
\begin{minipage}{0.8\linewidth}
\subfloat[Nine anchors centred]{
\includegraphics[width=0.45\linewidth]{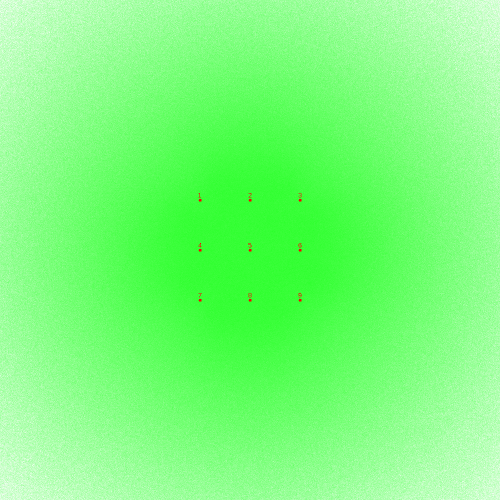}
\label{fig:mle91}
}\quad
\subfloat[Four anchors on the edges]{
\includegraphics[width=0.45\linewidth]{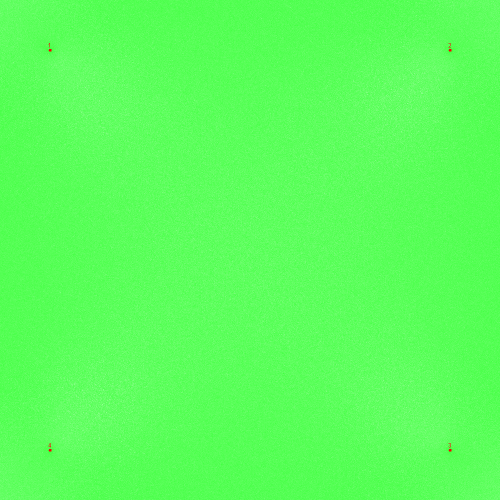}
\label{fig:mle41}
}\\
\subfloat[Five anchors in more challenging positions]{
\includegraphics[width=0.45\linewidth]{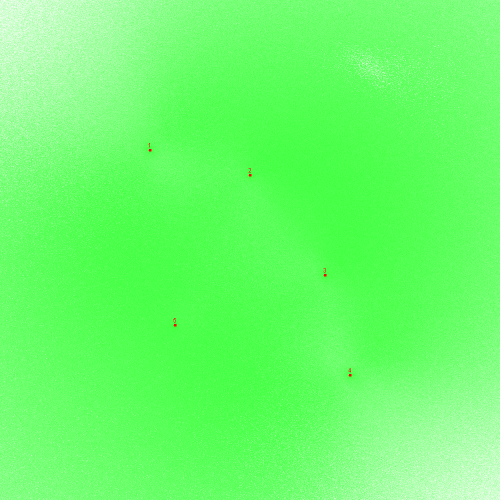}
\label{fig:mle51}
}\quad
\subfloat[Four anchors on the edges (40\% NLOS probability)]{
\includegraphics[width=0.45\linewidth]{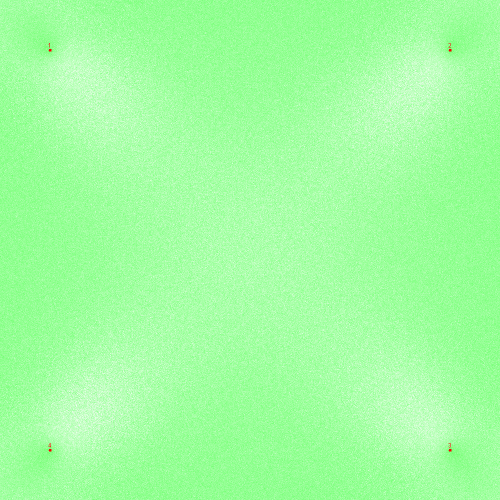}
\label{fig:mle44}
}
\caption[Performance of the MLE-$\Gamma$ algorithm. The algorithm has been fitted to the error distribution by drawing 100000 samples from the error model used in the simulation.]{}
\label{fig:mle}
\end{minipage}
\end{figure}

Figure~\ref{fig:mle} shows outstanding simulation results for the MLE-$\Gamma$ algorithm. Unless otherwise noted, Figures~\ref{fig:mle91}-\ref{fig:mle44} use the same simulation parameters than in the preceding simulations. The simulation results are hardly to distinguish and show a very homogeneous spatial distribution for the whole simulation area. Even if the probability for the occurrence of NLOS errors is raised to 40\% (as shown in Figure~\ref{fig:mle44}), the average result and the spatial distribution of the position error is not changing very much. For the MLE-$\mathcal{N}$ algorithm the simulation results are quite similar to the MLE-$\Gamma$ results, only a little weaker and because of that not illustrated here.

\newpage
\section{real world evaluation}
\label{deployment}

In order to measure the effectiveness of the six algorithms introduced in Section~\ref{relatedwork} and NLLS with real sensor network data and to be able to compare the results with the executed simulations, we recorded the data of a series of different test runs. The experiments were carried out using a modified version of the Modular Sensor Board (MSB) A2 \citep{TR-B-08-15} node which is equipped with a Nanotron nanoPAN 5375 \citep{nanopan} transceiver. This hardware enables the sensor nodes to measure inter-node ranges using \ac{TOF} in the 2.4 GHz frequency band. The experiments took place on the second floor of our Computer Science Department during daytime.

\begin{figure}[htb]
  \centering
  \includegraphics[width=.95\linewidth]{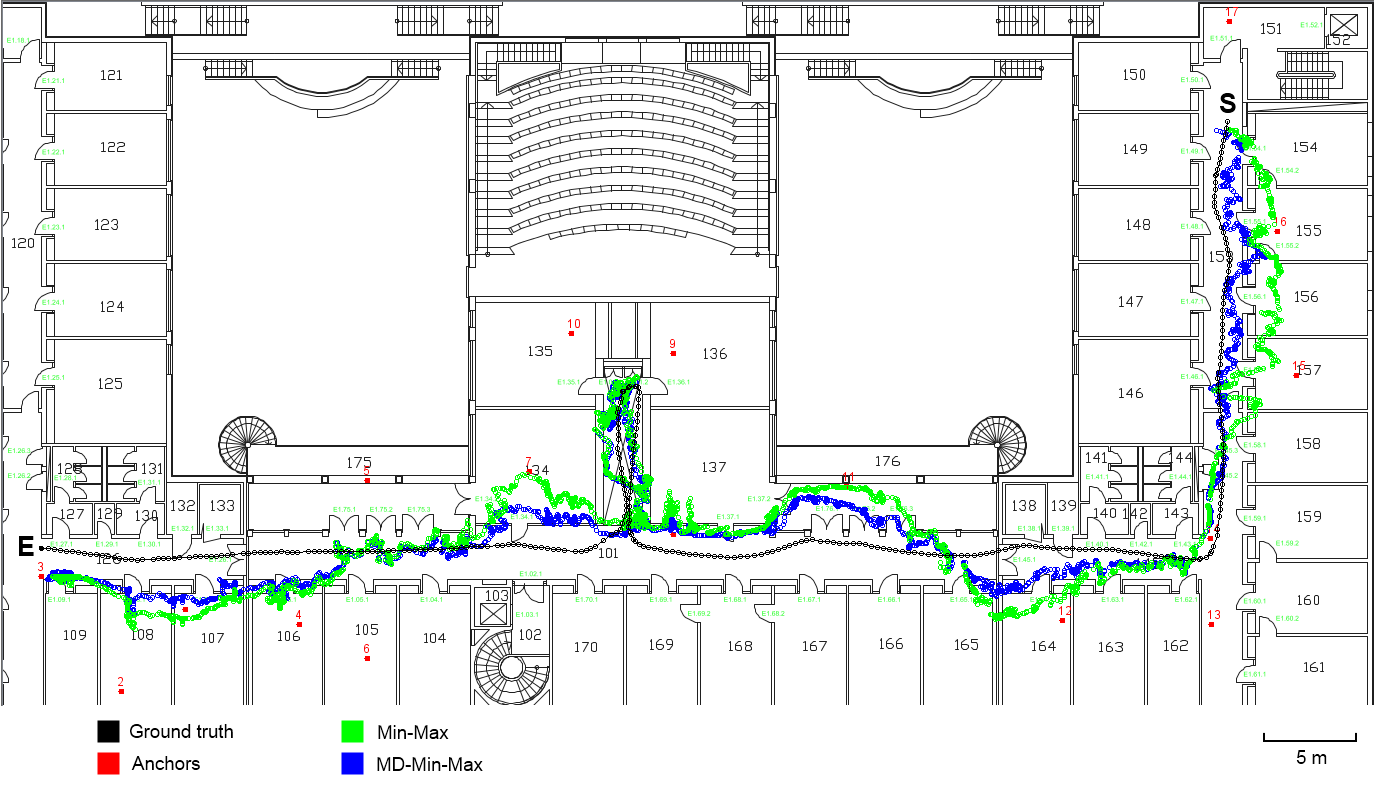}
  \caption[Position estimates on the second floor of our Computer Science Department. The length and the width of the experiment area is about 80~m $\times$ 40~m.]{}
  \label{fig:real_world_run2}
\end{figure}

Figure \ref{fig:real_world_run2} shows one exemplary campaign of measurements following a route among offices, laboratories and with a few people walking around. For the reason of clarity, we plotted only the results of Min-Max and \ac{MD-Min-Max} using a Kalman filter. The starting point is denoted by ``S'', the endpoint is denoted by ``E'' and the total length of the path was about 100 meters. In each run, we used 17 anchors which were deployed throughout the building. Most of the anchors were placed in office rooms with doors closed. Only a small fraction of nodes was placed on the hallway, in case of Figure~\ref{fig:real_world_run2}, there were four nodes. Ground truth was measured with the aid of a robot system developed at our Department using a Microsoft Kinect. This reference system provides about 10~cm positioning accuracy. The robot carried the unlocalised node and followed a predefined path with a predefined speed. We used the maximum movement speed of the robot, which is 0.5~m/s. In total, we performed over 5300 localisations when adding up all test runs. The nanoPAN achieves ranging precision of around 2.85 m in average and the RMSE is 4.32~m. However, the ranging error can be as large as 20~m. We even encountered measurement errors up to 75~m in rare cases.

The quantitative results of the seven localisation algorithms are shown in Table \ref{table_loc_results}. The average anchor degree throughout all experiments was 7.48. Additionally, Table \ref{table_loc_results} contains the results of multilateration using NLLS to give a comparison to a well known general purpose algorithm. As it can be seen, \ac{MD-Min-Max} outperforms the other algorithms in terms of localisation accuracy with achieving an average error of 1.63~m. The basic Min-Max algorithm (2.05~m) is still more than twice as good as NLLS (4.43~m) which serves as a reference algorithm. The good performance of Min-Max (and therefore also the other Min-Max algorithms) is not surprising because the inter-anchor distances were relative short (between 5 and 10 meters) and the mobile node took mainly positions within the bounds of the network. As we know from Section \ref{section_spatial_evaluation} this is the optimal situation for Min-Max algorithm. This fact is also observed by \citet{savvides2002} and proved by \citet{langendoen2003}. All three enhanced Min-Max algorithms outperform the original one: \ac{E-Min-Max} (W2) (1.46\%), \ac{E-Min-Max} (W4) (4.39\%) and \ac{MD-Min-Max} (20.48\%). Furthermore, all Min-Max based algorithms show quite small maximum positioning error as they bound the estimate inside the \ac{IR}.

To make NLLS, MLE-$\mathcal{N}$ and MLE-$\Gamma$ comparable, they all use the same three initial starting points for their optimization procedure. MLE-$\mathcal{N}$ outperforms NLLS by 52.82\% and MLE-$\Gamma$ even by 56.43\%. NLLS neither considers the real distribution of the error nor bounds the estimate and thus can be easily misled by the \ac{NLOS} error. Both MLE algorithms perform significantly better due to the consideration of the positive bias of the real error. MLE-$\Gamma$ achieves even a better average error (1.93~m) than the best \ac{E-Min-Max} algorithm (1.96~m). However, the outstanding results shown in Section~\ref{section_spatial_evaluation} cannot be reproduced in real world experiments. The reason for this lies in the error model of the simulation. Even if the error model of the simulation and the measured error of the real world experiment nearly have the same distribution, the origin of this distribution is quite different.

In a real world experiment, the occurrence of a certain error is twofold: while the measurement error part could be described as random noise with a certain distribution, the reflection and multipath error part is not that random. This path error is linked to the position in the building and can be characterised by the real position and the properties of the physical structure of the building. If the experiment is a result of a walk through the building none of two parameters could be characterized as random, even not if we use a random walk movement model. If we are at a certain position $P_n$ the next measured position $P_{n+1}$ depends on $P_n$ if the movement speed is limited. The physical structure also directly depends on the position. The error model of the simulation does not take this into account and realizes the error distribution in a total random order without any position dependencies. This results in a very different behaviour from simulation and real world 
experiments. To overcome this, ray-tracing error models and structure maps can be used. Alternatively, if we adopt the algorithm to the dependencies in the distribution, we would end up in zero position error and hard-coding the whole experiment with its exact path into the algorithm, which would be a useless approach of course.

\begin{figure}[htb]
\centering
\includegraphics[width=.95\linewidth]{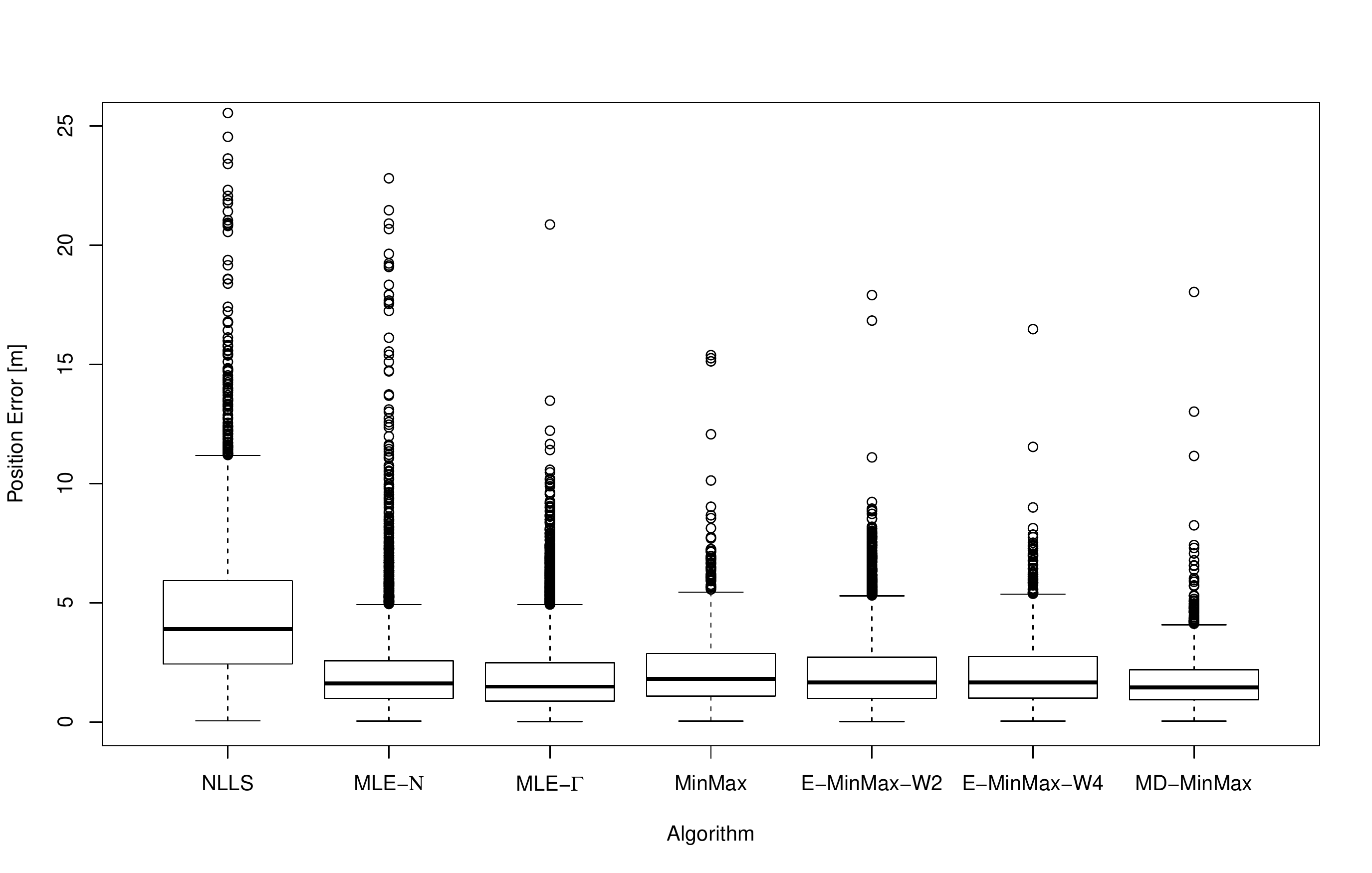}
 \caption[The distribution of the position error for the selected algorithms. Outliers are cropped at 25~m for better readability.]{}
 \label{fig:real_world_histogram_boxplot}
\end{figure}

The fact that the RMSE of Min-Max, \ac{E-Min-Max} (W2), \ac{E-Min-Max} (W4), \ac{MD-Min-Max}, and both \ac{ML} estimators is much smaller than the RMSE of the distance measurements tells us that these algorithms performed very well relative to the quality of the distance measurements available. NLLS with having a RMSE only slightly larger than the RMSE of the distance measurements showed acceptable performance. The distribution of the localisation error of all algorithms is shown in Figure~\ref{fig:real_world_histogram_boxplot} where the vertical axis is the localisation error in meters and the horizontal axis is the corresponding algorithm. NLLS shows poor performance compared to the other algorithms. Also the RMSE is much larger than that of the other algorithms. \ac{MD-Min-Max} has the smallest spread among all algorithms when regarding only non-outliers and also the lowest median of the error. Furthermore, the interquartile
range of \ac{MD-Min-Max} is the smallest among all algorithms. This algorithm outperforms even \ac{E-Min-Max} (W4) by more than 16\%. This performance gain is mainly achieved by adjusting the parameters of the algorithm to the error distribution (see Figure~\ref{fig:histogram_ranging_and_fitting}) of the used distance measurement hardware as described in Section~\ref{section:md-min-max}. Only the \ac{ML} estimators achieve a comparable performance gain but they suffer from many large outliers compared to the Min-Max algorithms. \ac{E-Min-Max} (W2) and \ac{E-Min-Max} (W4) show nearly the same performance. \ac{E-Min-Max} (W4) is slightly better because its weighting function is optimized for locations inside the perimeter of the anchors as was mostly the case.

Note, that \ac{MD-Min-Max} is quite sensitive to the parameters of the membership function. When assuming a Gaussian error distribution on our statistical data and using the three-sigma rule, the membership function is characterized by $[-8.3; 2.4; 13.1]$. With this function, the average error regresses to 1.89 meters. \ac{MD-Min-Max} can even become the worst algorithm, when the membership function does not fit the data. For example, choosing $[6; 12; 18]$ for the membership function will result in an average error of 2.19 meters. A careful analysis of the statistical data is necessary for good results. The membership function is characteristic to a deployment, e.g. a building or one of its floors, and results are of similar quality for multiple runs in such a deployment.

Obviously, the position accuracy could be improved using some filtering techniques, such as Kalman or particle filters, but the aim of this paper is to show and compare the performance of the used localisation algorithms without using any of those filtering techniques.

\newpage
\section{Conclusion}
\label{conclusion}

We have presented the \ac{MD-Min-Max} algorithm as an optimization of the Min-Max and also both \ac{E-Min-Max} algorithms. We have shown that performance can be improved noticeably in most scenarios, if simple assumptions about the error distribution are regarded by the algorithm. Although modelling the extremely arbitrary indoor distance measurement error is hard, experimental results imply that our method fits the real world positioning effectively and efficiently.

This behaviour is corroborated by the simulations of the spatial position error and validated by the experiments conducted where the accuracy improvement ranged from 15.5\% to 20.5\% (disregarding NLLS). While MLE-$\mathcal{N}$ and MLE-$\Gamma$ improve the performance of NLLS by over 50\%, they cannot attain the high accuracy of \ac{MD-Min-Max} completely. In particular the larger maximum error is a disadvantage when compared to \ac{MD-Min-Max}, which is especially important for practical applications.

Another disadvantage of MLE-$\mathcal{N}$ and especially MLE-$\Gamma$ is their high computational cost, whereas \ac{MD-Min-Max} is lightweight and can be computed on the same hardware as the \ac{E-Min-Max} algorithms. Thus, it is a good choice for the localisation in \acp{WSN} and for cooperative localisation scenarios, where every node has to compute its own position often and fast.

The analysis of the spatial error distribution shows, that \ac{E-Min-Max}~(W2) has the lowest average error in the simulation but does not reach lower errors in many real world experiments because often the high accuracy is achieved by a very good performance outside of the anchor hull which is often not of interest for indoor deployments. The basic Min-Max algorithm has the highest average error but shows good results in practical experiments because the areas with low errors are located as a continuous shape inside the convex hull of the anchors. Most real world anchor set-ups have a similar scenario because commonly anchors are placed near walls and not in the middle of rooms. Due to this observations \ac{E-Min-Max}~(W4) and \ac{MD-Min-Max} perform very good in most real world deployments because they have a lower worst case error and their low error regions are also very large and continuous. The visualization of the spatial error distribution also shows that a combination of \ac{E-Min-Max} (W2) and \ac{E-Min-Max}~(W4) would be a good approach to get more precision without any assumptions about the distance error distribution on which \ac{MD-Min-Max} relies.

Future work should address more optimization regarding the spatial error distribution. We have shown that the optimal choice of an algorithm at each point in time would provide even better localisation results. It should also be possible to integrate a weighting component based on the distance error distribution into other more complex algorithms to gain performance improvements.


\begin{acronym}\addtolength{\itemsep}{-0.5\baselineskip}
 \acro{E-Min-Max}[E-Min-Max]{Extended Min-Max}
 \acro{MD-Min-Max}[MD-Min-Max]{Membership Degree Min-Max}
 \acro{MF}{Membership Function}
 \acro{WSN}[WSN]{Wireless Sensor Network}
 \acro{TOF}[TOF]{time-of-flight}
 \acro{IR}[$\mathit{IR}$]{intersection region}
 \acro{NLOS}[NLOS]{non-line-of-sight}
 \acro{ML}[ML]{maximum likelihood}
 \acro{MLE}[MLE]{maximum likelihood estimator}
 \acro{pdf}[p.d.f.]{probability density function}
\end{acronym}


\newpage
\bibliographystyle{plainnat}
\bibliography{../bibtex/bibliography}

\newpage
\renewcommand*\listfigurename{Figure captions}
\newlength{\fig}
\settowidth{\fig}{Figure\,99:~}
{\renewcommand*\numberline[1]{\llap{\makebox[\fig][l]{Figure\,#1:~}}}
\makeatletter
\renewcommand*\l@figure[2]{\leftskip\fig\noindent#1\par}
\makeatother
\listoffigures}

\newpage
\section*{Tables}

\begin{table}[h]
\centering
\caption{Quantitative results for the localisation task}
\label{table_loc_results}
\begin{tabular}{|l|r|r|r|}
  \hline
  Algorithm & MAE [m] & RMSE [m] & MAX [m] \\
  \hline\hline
  NLLS                           			& 4.43 & 5.24 & 25.55 \\
  MLE-$\mathcal{N}(2.43,3.57^2)$  	& 2.09 & 2.84 & 22.81 \\
  MLE-$\Gamma(3.3,0.58)$		& 1.93 & 2.52 & 27.04 \\
  Min-Max             				& 2.05 & 2.42 & 15.39 \\
  \ac{E-Min-Max} (W2) 			& 2.02 & 2.49 & 17.91 \\
  \ac{E-Min-Max} (W4) 			& 1.96 & 2.34 & 16.48 \\
  \ac{MD-Min-Max} (MF by M2)		& 1.63 & 1.89 & 18.04 \\
  \hline
\end{tabular}
\end{table}

\newpage
\section*{Figures}

\renewcommand{\figurename}{Figure}

\end{document}